\documentclass[fleqn,usenatbib]{mnras}

\usepackage[T1]{fontenc}
\usepackage{ae,aecompl}
\usepackage{tabularx}

\usepackage{epsf,palatino,url,wrapfig,array,setspace,amsmath,amssymb,fancyhdr,multirow,lscape,appendix,rotating}
\usepackage{graphics,epsfig,txfonts}
\usepackage{graphicx}

\usepackage{longtable}
\usepackage{amssymb}
\usepackage{xcolor}
\usepackage{color}
\usepackage{lipsum}
\usepackage{textcomp}
\usepackage{threeparttable}
\usepackage{subcaption}
\newsavebox{\measurebox}

\usepackage{enumerate}
\usepackage{placeins}
\usepackage{float}
\usepackage[normalem]{ulem}
\usepackage{soul}
\usepackage{blindtext}

\usepackage{tikz,hyperref}
\definecolor{lime}{HTML}{A6CE39}
\DeclareRobustCommand{\orcidicon}{%
	\begin{tikzpicture}
	\draw[lime, fill=lime] (0,0) 
	circle [radius=0.16] 
	node[white] {{\fontfamily{qag}\selectfont \tiny ID}};
	\draw[white, fill=white] (-0.0625,0.095) 
	circle [radius=0.007];
	\end{tikzpicture}
	\hspace{-2mm}
}

\foreach \x in {A, ..., Z}{%
	\expandafter\xdef\csname orcid\x\endcsname{\noexpand\href{https://orcid.org/\csname orcidauthor\x\endcsname}{\noexpand\orcidicon}}
}

\newcommand{\orcid}[1]{\href{https://orcid.org/#1}{\textcolor[HTML]{A6CE39}{\orcidicon}}}

\newcommand{\nicer}{\textit{NICER}\xspace}

\newcommand{\ero}{{\it eROSITA}\xspace}
\newcommand{\rosat}{{\it ROSAT}\xspace}

\newcommand{\xmm}{{\it XMM-Newton}\xspace}

\newcommand{\swift}{{\it Swift}\xspace}

\newcommand{\oergcm}[1]{$10^{#1}$ erg cm$^{-2}$ s$^{-1}$}
\newcommand{\ergs}[1]{$\times 10^{#1}$ erg s$^{-1}$}
\newcommand{\oergs}[1]{$10^{#1}$ erg s$^{-1}$}
\newcommand{\Halpha}{{H${\alpha}$}\xspace}

\newcommand{\lxp}{LXP\,69.5\xspace}
\newcommand{\rxj}{RX\,J0529.8$-$6556\xspace}

\newcommand{\eqb}{\begin{eqnarray}}
\newcommand{\eqe}{\end{eqnarray}}

\graphicspath{{./}{finalplots/}}

\title[LXP 69.5: the 2020 outburst]{RX J0529.8$-$6556: a BeXRB pulsar with an evolving optical period and out of phase X-ray outbursts}

\author[H. Treiber et al.]
{H. Treiber\orcid{0000-0003-0660-9776},$^{1,2}$\thanks{E-mail: htreiber22@amherst.edu}
G.~Vasilopoulos\orcid{0000-0003-3902-3915},$^1$\thanks{E-mail: georgios.vasilopoulos@yale.edu}
C. D. Bailyn,$^1$
F. Haberl\orcid{0000-0002-0107-5237},$^3$
K.~C.~Gendreau,$^{4}$
\newauthor
P.~S.~Ray\orcid{0000-0002-5297-5278},$^{5}$
C. Maitra\orcid{0000-0002-0766-7313},$^3$
P. Maggi\orcid{0000-0001-5612-5185},$^6$
G.~K. Jaisawal\orcid{0000-0002-6789-2723},$^7$
A. Udalski\orcid{0000-0001-5207-5619},$^8$
J. Wilms,$^9$
\newauthor
I.~M. Monageng\orcid{0000-0002-4754-3526},$^{10,11}$ 
D.~A.~H. Buckley\orcid{0000-0002-7004-9956},$^{10}$ 
O.~ König,$^9$
S. Carpano$^3$
\\
$^1$Department of Astronomy, Yale University, PO Box 208101, New Haven, CT 06520-8101, USA\\
$^2$ Department of Physics and Astronomy, Amherst College, C025 New Science Center, 25 East Dr., Amherst, MA 01002-5000, USA\\
$^3$Max-Planck-Institut f\"ur extraterrestrische Physik,Giessenbachstra{\ss}e, 85748 Garching, Germany\\
$^4$X-Ray Astrophysics Laboratory, NASA Goddard Space Flight Center, Greenbelt, MD 20771, USA\\
$^5$Space Science Division, U.S. Naval Research Laboratory, Washington, DC 20375, USA\\
$^6$Université de Strasbourg, CNRS, Observatoire astronomique de Strasbourg, UMR 7550, 67000, Strasbourg, France\\
$^7$National Space Institute, Technical University of Denmark, Elektrovej 327-328, DK-2800 Lyngby, Denmark\\
$^8$Astronomical Observatory,
University of Warsaw, Al. Ujazdowskie 4, 00-478, Warszawa, Poland\\
$^9$ Dr. Karl Remeis-Observatory and Erlangen Centre for Astroparticle Physics, Sternwartstr. 7, 96049 Bamberg, Germany\\
$^{10}$ South African Astronomical Observatory, P.O. Box 9, Observatory, Cape Town 7935, South Africa\\
$^{11}$ Department of Astronomy, University of Cape Town, Private Bag X3, 7701 Rondebosch, South Africa\\
}

\date{Accepted XXX. Received YYY; in original form ZZZ}

\pubyear{2021}

\begin{document}
\label{firstpage}
\pagerange{\pageref{firstpage}--\pageref{lastpage}}
\maketitle

\begin{abstract}

We report the results of \ero and \nicer observations of the June 2020 outburst of the Be/X-ray binary pulsar \rxj in the Large Magellanic Cloud, along with the analysis of archival X-ray and optical data from this source. We find two anomalous features in the system's behavior. First, the pulse profile observed by \nicer during maximum luminosity is similar to that observed by \xmm in 2000, despite the fact that the X-ray luminosity was different by two orders of magnitude. By contrast, a modest decrease in luminosity in the 2020 observations generated a significant change in pulse profile. 
Second, we find that the historical optical outbursts are not strictly periodic, as would be expected if the outbursts were triggered by periastron passage, as is generally assumed. 
The optical peaks are also not coincident with the X-ray outbursts. We suggest that this behavior may result from a misalignment of the Be star disc and the orbital plane, which might cause changes in the timing of the passage of the neutron star through the disc as it precesses. We conclude that the orbital period of the source remains unclear.
\end{abstract}




\section{Introduction}
High-mass X-ray binaries (HMXBs) consist of a compact object (i.e., a white dwarf, neutron star, or black hole) and an O- or B-star. 
A sub-class of HMXBs are Be/X-ray binaries \citep[BeXRBs; see][for a review]{2011Ap&SS.332....1R}, for which the donor star is a Be star surrounded by a variable decretion disc formed by ejected ionized gas. The Be disc is the source of emission lines and infrared excess that set Be stars apart from ordinary O/B-stars.
BeXRBs host the majority of X-ray pulsars, in which case the compact object is a young magnetized neutron star (NS). 
Material is accreted onto the NS by following the magnetic field lines and forming hotspots on the magnetic poles. Because of the misalignment of the NS magnetic and rotation axes, the orientation of the hotspots with respect to the observer changes according to the phase of the NS spin period, causing the observed modulations.    

Most BeXRBs have elliptical orbits.
During periastron passage, the NS comes close to the Be disc (or even passes through it), enabling mass transfer and the formation of a transient accretion disc around the NS. Material is eventually deposited onto the NS magnetic poles via magnetospheric accretion \citep{1973ApJ...179..585D}. 
Optical monitoring of BeXRBs can reveal periodic orbital modulation related to the orbital period, but also may show long-term variability due to Be disc truncation and precession effects \citep{2011MNRAS.416.2827M}.
X-ray and optical outbursts of BeXRBs usually coincide.
Type I X-ray outbursts occur periodically or quasi-periodically due to periastron passage mass transfer \citep{2013PASJ...65...41O}. Type II outbursts, on the other hand, happen irregularly, have longer durations, and are characterized by luminosities close or above the Eddington limit of the compact object. These rare events have been proposed to occur when a highly misaligned decretion disc around the Be star becomes eccentric, causing high mass transfer episodes \citep[][]{2014ApJ...790L..34M}.
During these major outbursts, BeXRBs account for some of the brightest accretion-driven stellar mass X-ray sources with luminosities that exceed \oergs{39} \citep{2018ApJ...863....9W,2020MNRAS.494.5350V}. 

As young binary systems, HMXBs are found in star-forming regions. 
Thus, the Small and Large Magellanic Clouds (SMC and LMC) are prime locations for HMXBs. The dwarf galaxies have well-determined distances of $\sim$50 kpc \citep[LMC,][]{2019Natur.567..200P} and $\sim$62 kpc \citep[SMC,][]{2014ApJ...780...59G}. Moreover, spectral studies are not affected by high foreground absorption since the Magellanic Clouds (MCs) are located away from the Galactic plane.
The SMC hosts more than 100 HMXBs, and all but one of the more than 60 confirmed pulsars are in BeXRBs \citep{2016A&A...586A..81H}. 
On the other hand, about $\sim$60 HMXBs candidates have been identified in the LMC, but pulsations have only been detected in about 20 of them. Despite the significant number of LMC systems, the study of the HMXB population is still incomplete given the large angular size of the galaxy and an insufficient coverage at energies above $\sim$2 keV. 
Among the LMC HMXBs, there is a high fraction of supergiant systems \citep{2019MNRAS.490.5494M,2018MNRAS.475.3253V,2016MNRAS.459..528A}. Two of these have been suggested to be probable Supergiant Fast X-ray Transients (SFXTs) based on their fast flaring properties \citep[see][]{2018MNRAS.475..220V,2021A&A...647A...8M}. 

The combined efforts of current X-ray observatories can bridge the gap between the study of the populations of HMXBs in the SMC and LMC. In particular, \ero's all-sky survey and the complementary capabilities of \nicer will identify and characterize more HMXBs in the LMC in the coming years. \ero is the primary instrument on board the Russian-German "Spectrum-Roentgen-Gamma" (SRG) mission \citep{2020arXiv201003477P}.
Using its seven identical X-ray telescopes, \ero's primary mission is to perform an imaging all-sky survey within four years. \ero will scan the sky eight times. While most of the sky is scanned every six months with an average exposure of $\lesssim$250 s,
the regions close to the ecliptic poles, where the great circles traced by the telescope on the sky intersect, are scanned six times per day for up to several weeks at a time. Since the LMC is located near the South Ecliptic Pole, it is an ideal laboratory for studying new X-ray outbursts, which can be monitored for weeks by \ero. 
Thus, \ero will help discover and characterize transients in the Magellanic Clouds \citep[e.g.,][]{2020ATel13828....1H,2020ATel13789....1H,2020ATel13844....1R}. 
To complement \ero, \nicer can follow-up on new X-ray transients within hours, leveraging a large effective area, superb timing resolution, and fast pointing capabilities. The \nicer X-ray Timing Instrument \citep[XTI,][]{2012SPIE.8443E..13G,2016SPIE.9905E..1HG} is a non-imaging, soft X-ray telescope aboard the \textit{International Space Station} (\textit{ISS}). 
Especially in the LMC, the combination of \ero and \nicer provides a new opportunity in the study of X-ray transients.

Data from other wavelength regimes also help to characterize HMXBs.
In particular, long-term optical behavior can provide insights on the nature of donor stars (and the decretion disk) in BeXRBs. For the LMC and SMC, up to 20 years of monitoring data can be provided by the Optical Gravitational Lensing Experiment (OGLE)  \citep{1992AcA....42..253U}.

In this paper we present the X-ray and optical properties of a known but poorly studied system in the LMC, the BeXRB pulsar \rxj.
The source, which is also known as \lxp, was first discovered with \rosat as a transient in the LMC in 1993 \citep{1997A&A...318..490H}. They established the system as a BeXRB with a likely pulsation period of 69.5 seconds.
In Sect. \ref{sec:2020xray} we present the spectral and temporal properties of \lxp based on the analysis of archival and newly obtained X-ray data. Section \ref{sec:ogle} includes the investigation of the long-term optical light curve of the system. We discover an evolving optical period and establish that the optical and X-ray flares do not coincide. With ten years of OGLE monitoring in V and I bands, we model the quasi-periodicity and show the system gets redder as it brightens. A SALT optical spectrum of the donor star is presented in Sect. \ref{sec:salt}, showing a double-peaked H$\alpha$ emission profile. Finally, in Sect. \ref{sec:disc}, we estimate the NS magnetic field strength and discuss the observational properties in the context of the system's geometry.

\section{X-ray properties}
\label{sec:2020xray}

In the following sections we provide details about the analysis of the X-ray data collected during the 2020 X-ray outburst, as well as the derived X-ray spectral and temporal properties of the system. We also investigate the long-term variability of the system using archival X-ray data (i.e., prior to the 2020 outburst) obtained by \swift and \xmm. 

\subsection{The 2020 X-ray outburst}

In the course of the first all-sky survey (eRASS1), the \ero instrument on board the Russian/German SRG mission detected a strong outburst from \lxp \citep{2020ATel13828....1H}.
After an initial period of 20 days at the start of the survey when the source was not detected, \ero began scanning the location of the system again from 2020-06-06 00:58 UTC (MJD 59006.04) for about 30-40 s every four hours and found \lxp in outburst.

Following the SRG/\ero discovery of a strong outburst from \lxp, we triggered \nicer ToO observations to investigate the temporal properties (i.e., pulse profile shape) and spectral properties while comparing with the \ero data. 

\subsubsection{eROSITA data}

The data were reduced with a  pipeline based on the \ero Standard Analysis Software System \citep[eSASS,][]{2020arXiv201003477P},
which determines good time intervals, corrupted events and frames, dead times, masks bad pixels, and applies pattern recognition and energy calibration. Star-tracker and gyro data were used to assign celestial coordinates to each reconstructed X-ray photon.

\begin{figure}
    \centering
    \includegraphics[width=1.0\columnwidth]{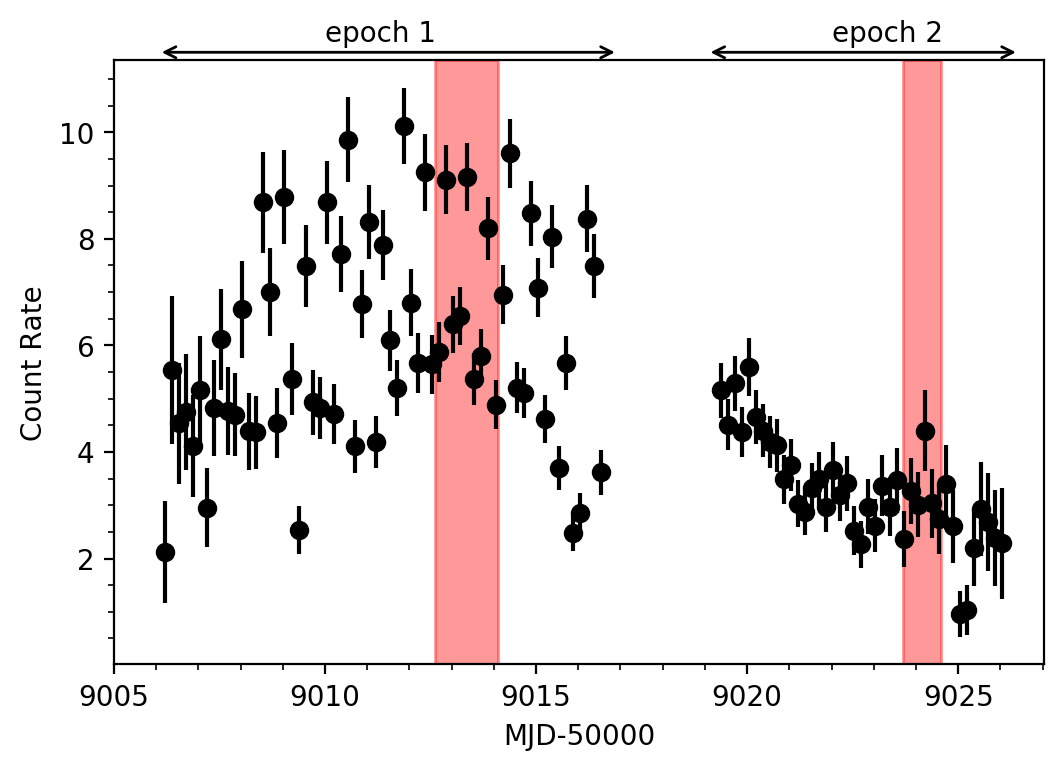}
    \caption{eRASS light curve of \lxp showing day-to-day variations of the source. Each point typically uses 30-40\,s exposure. \nicer observed the source during the two epochs shown by the red shading.}
    \label{fig:erosita_lc}
\end{figure}

To create an X-ray light curve we used data from all seven telescope modules.
Source detection was performed simultaneously on all the images in the standard \ero energy bands of 0.2--0.6\,keV, 0.6--2.3\,keV, and 2.3--5.0\,keV.
The vignetting and point spread function (PSF)-corrected count rates averaged over a single scan in the energy range 0.3--10.0 keV are displayed in Fig. \ref{fig:erosita_lc}. 
Monitoring of the source with \ero lasted for about 20 days, with a two day gap around the middle of the observing window. 
The temporal properties of \lxp seem to change between the two epochs. At the start of the monitoring (i.e., epoch 1 or MJD 59005--59017), \ero data show higher variability compared to the second epoch (MJD 59018--59027), even though the count rate decreased by a small factor in that time.

For the spectral analysis only the telescope modules with on-chip filter are used (i.e., ~TM1,TM2, TM3, TM4, and TM6). For spectral extraction, circular regions with a radius of 32\arcsec\ and 50\arcsec\ were used for the source and background extraction, respectively. Given the change in the temporal properties of \lxp as identified in the X-ray light curve we extracted separate spectra for epochs 1 and 2 to investigate if the different epochs also show a characteristic change in the spectral properties (see Section \ref{sec:spectra}).

\subsubsection{NICER data}
\label{sec:nicer_data}

The \nicer X-ray Timing Instrument \citep[XTI,][]{2012SPIE.8443E..13G,2016SPIE.9905E..1HG} is a non-imaging, soft X-ray telescope aboard the \textit{International Space Station} (ISS). The XTI consists of an array of 56 co-aligned concentrator optics, each associated with a silicon drift detector \citep{2012SPIE.8453E..18P}, operating in the 0.2--12 keV band. The XTI provides high time resolution ($\sim$100 ns) and spectral  resolution of $\sim$85 eV at 1 keV. It has a field of view of $\sim$30 arcmin$^2$ in the sky and an effective area of $\sim$1900 cm$^2$ at 1.5 keV (with the 52 currently active detectors). 

For the current study we analysed \nicer data obtained between MJD  59012 and 59025. Most \nicer snapshots were clustered around two epochs as marked in Fig. \ref{fig:erosita_lc}. These epochs are contained within the \ero epochs so we will refer to them similarly as \nicer epochs 1 and 2.
Data were reduced using {\tt HEASOFT} version 6.26.1, \nicer DAS version 2019-06-19\_V006a, and the calibration database (CALDB) version 20200315 (redistribution matrix and ancillary response files). 
Since \nicer is not an imaging instrument, an important step is the determination of the X-ray background during the observations. 
There are conditions for which the background models are ineffective, calling for more aggressive filtering. This step is particularly important for spectroscopic studies of faint systems, since the background contribution can be comparable to the source intensity. However, for timing analysis and period search, higher statistics are preferred.

For the timing analysis, we selected good time intervals according to the following space weather conditions: 
\textit{ISS} not in the South Atlantic Anomaly region, source elevation $>20^\circ$ above the Earth limb ($>30^\circ$ above the bright Earth), pointing offset $\le 54$ arcsec, and magnetic cutoff rigidity (\texttt{COR\_SAX}) > 1.5 GeV/c.
Barycentric corrections to time of arrivals were performed using the \texttt{barycorr} tool and the JPL DE405 planetary ephemeris.

For spectroscopy, we used more strict filtering criteria (i.e., source elevation $>15^\circ$), and screened the event list based on intensity in the 10--15 keV range.
For the X-ray background (BG) estimation two different and independent modeling tools have been developed.
Here we will provide a short description of the tools, but we note that a comprehensive explanation is available in the \nicer online portal\footnote{Models are publicly available: \url{https://heasarc.gsfc.nasa.gov/docs/nicer/tools/nicer_bkg_est_tools.html}}.
As the \textit{ISS} and \nicer travel through a wide range of geomagnetic latitudes -- each with its own BG characteristics -- individual observations have different BG levels that need to be understood. As a first step, a grid of \nicer blank-sky spectra corresponding to the blank-sky pointings of Rossi X-ray Timing Explorer \citep[see][]{2006ApJS..163..401J} is created.
The ``space weather'' method (Gendreau et al., in prep.) uses environmental data to parse the BG database.  The BG spectrum is generated by combining these blank-sky spectra weighted according to space weather conditions and magnetic cutoff rigidities common to both the pulsar and BG-fields observations.
The {\tt 3C50} tool (Remillard et al, in prep.) uses a number of BG proxies in the \nicer data to define the basis states of the BG database. These BG proxies include graded event data that discriminate some X-ray events landing near the center of each detector from those interacting near the detector edge under a field aperture. Another proxy is high energy X-ray events, which are unlikely to have been focused by the X-ray mirrors. The tool builds a predicted BG spectrum from similarly selected data from the BG database.

By implementing both methods we found that the environmental model yields a lower BG, which does not scale linearly with the {\tt 3C50} model. In fact, the latter predicts a BG which is a factor of 2--4 higher between 2--10 keV. 
We note that for temporal studies, background estimates do not affect any periodic signal, but could affect the baseline determination and estimates of pulsed fraction. For spectral studies, however, significant residuals will appear in any spectral model lacking a good background estimate. Because of the contemporaneous \ero covering, we may test which BG model is more representative by comparing the spectral properties between \ero and \nicer. As we will discuss in Sect. \ref{sec:spectra} the {\tt 3C50} model is more appropriate for the case of \lxp.

\subsection{Temporal properties}
\label{sec:nicer_time}

\begin{figure}
    \centering
    \includegraphics[width=1.0\columnwidth]{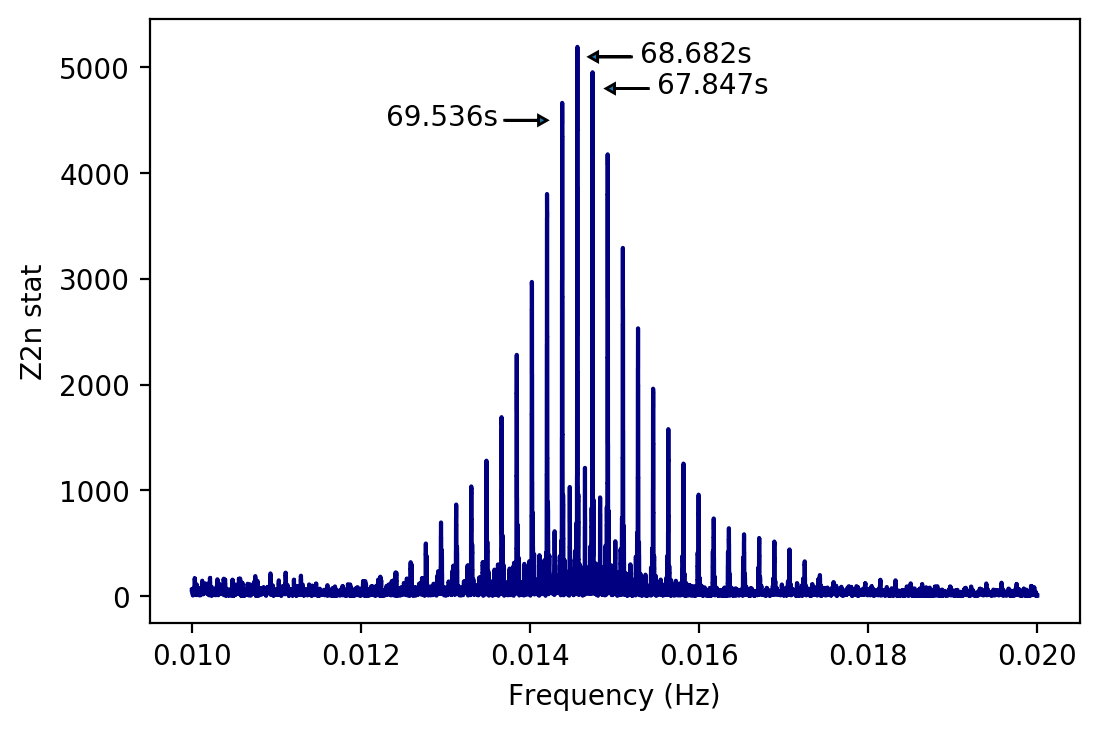}
    \caption{{\tt HENdrics} epoch-folding Zsearch between 0.01 and 0.02 Hz. Observational sampling led to multiple alias peaks. 
    }
    \label{fig:zsearch}
\end{figure}

\begin{table}
\caption{Pulse Timing Parameters for \lxp\label{tab:timing}} 
\begin{tabular}{ll}
\hline\hline
\multicolumn{2}{c}{Fit and data-set} \\
\hline
MJD range\dotfill & 59012.6--59014.1 \\ 
Number of TOAs $^{(a)}$\dotfill &  52\\
Rms timing residual (s)\dotfill & 1.87 \\
\hline
\hline
\multicolumn{2}{c}{Set Quantities} \\ 
\hline
Right ascension (J2000), $\alpha$ (hh:mm:ss)\dotfill & 5:29:47.84 \\ 
Declination (J2000), $\delta$ (dd:mm:ss)\dotfill & $-$65:56:43.7 \\ 
Epoch of frequency determination $t_{\mathrm 0}$ (MJD)\dotfill &  59013 \\ 
\hline
\multicolumn{2}{c}{Measured Quantities $^{(b)}$} \\ 
\hline
Pulse frequency, $\nu_{\rm 0}$ (s$^{-1}$)\dotfill &  0.0145593(2)  \\ 
First derivative of pulse frequency, $\dot{\nu}$ (s$^{-2}$)\dotfill &   1.8(7)$\times 10^{-11}$ \\ 
\hline
\multicolumn{2}{c}{Assumptions} \\
\hline
Solar system ephemeris model\dotfill & DE405 \\
Time units \dotfill &  TDB$^{(c)}$ \\
\hline
\end{tabular}\\
\footnotesize{Table notes: $^{(a)}$ Time of arrivals are calculated by phase folding events for 52 intervals and comparing them to a template pulse profile. $^{(b)}$ best fit values and 1$\sigma$ uncertainties. $^{(c)}$ Barycentric dynamic time.}
\end{table}

\begin{figure*}
    \centering
    \begin{minipage}[b]{.49\textwidth}
  \subfloat
    []
    {\label{fig:figA}\includegraphics[width=\textwidth]{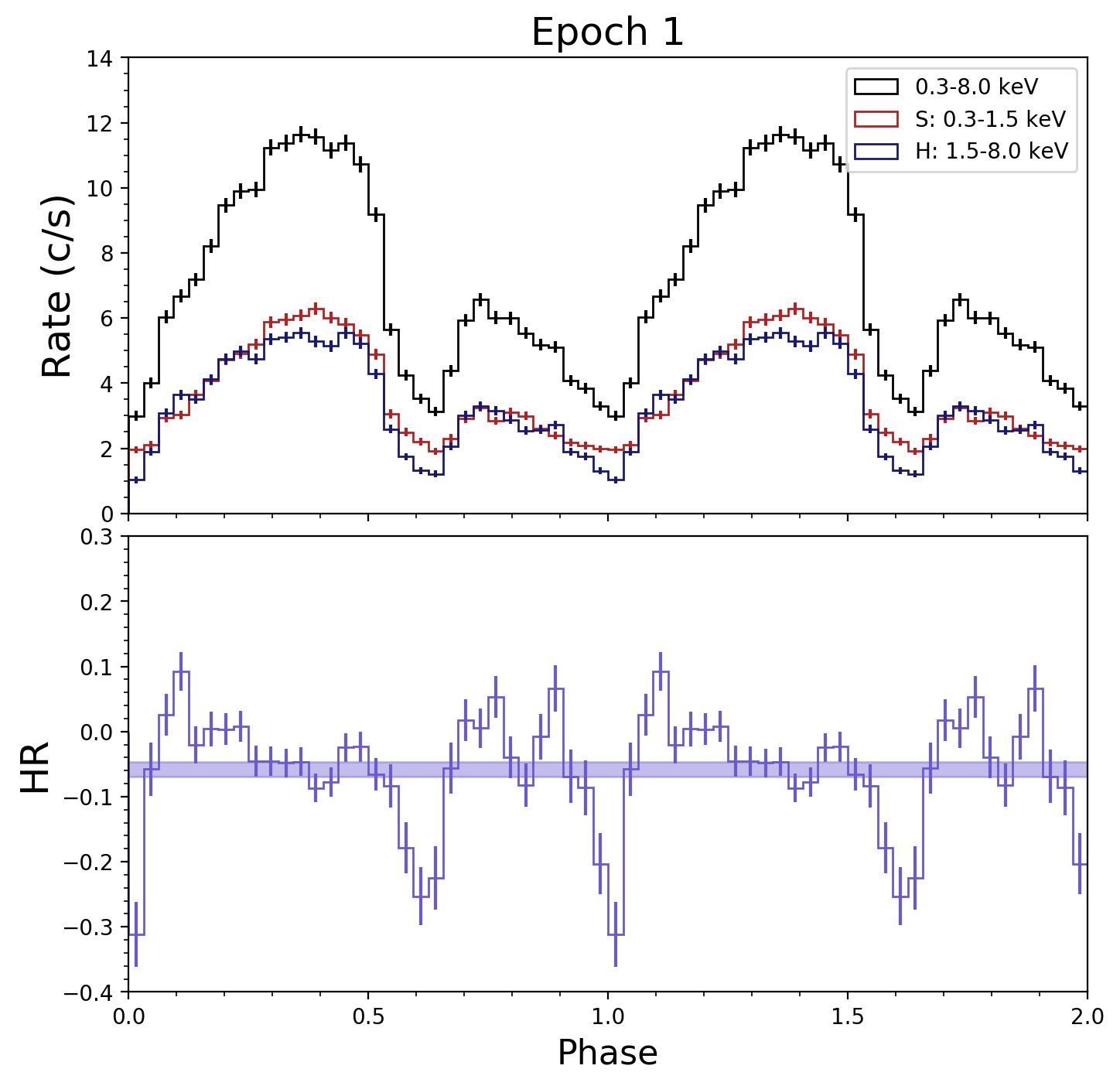}}
\end{minipage}
\begin{minipage}[b]{.49\textwidth}
  \subfloat
    []
    {\label{fig:figB}\includegraphics[width=\textwidth]{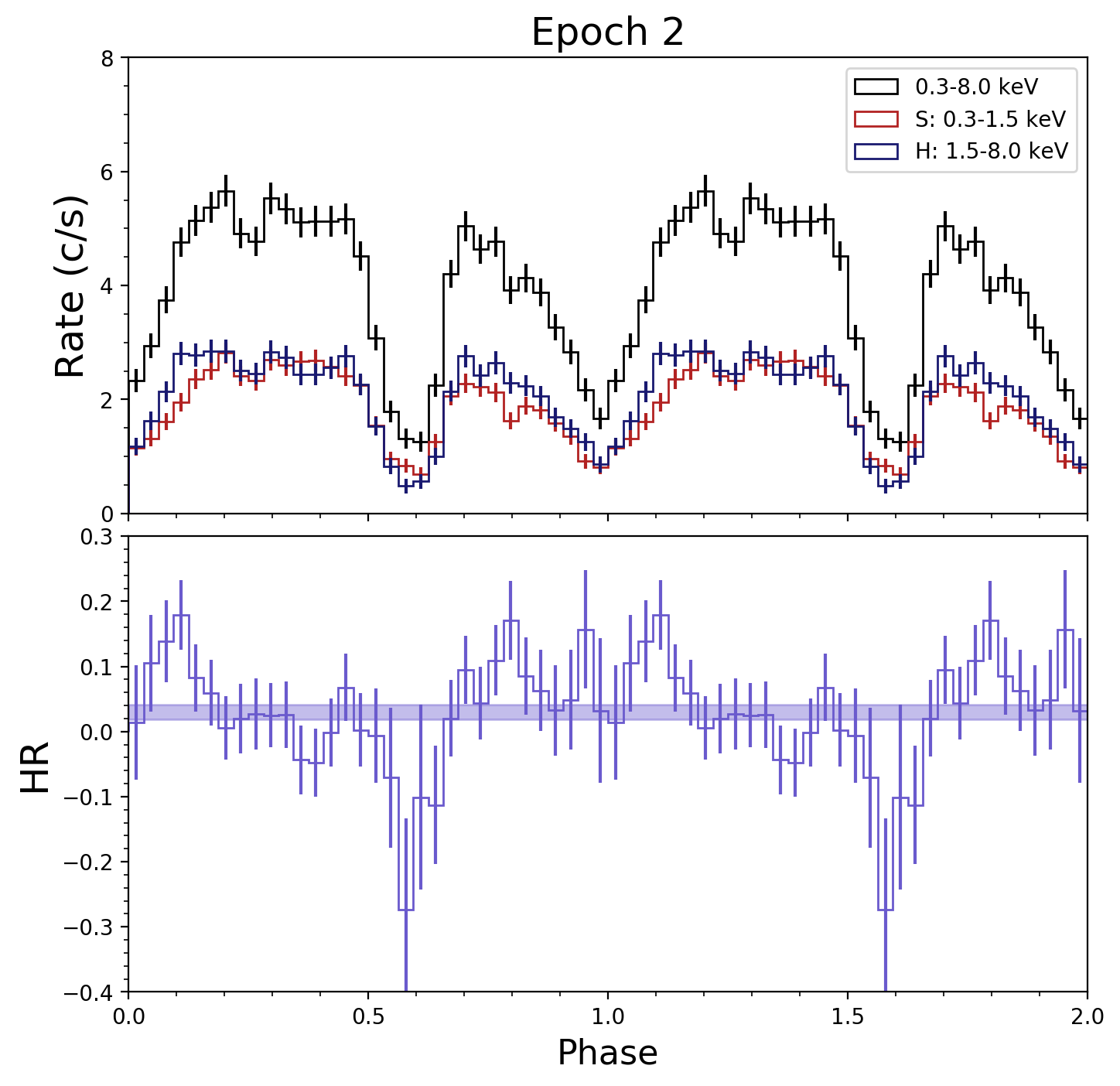}}
\end{minipage}
    \begin{minipage}[b]{.49\textwidth}
  \subfloat
    []
    {\label{fig:figC}\includegraphics[width=\textwidth]{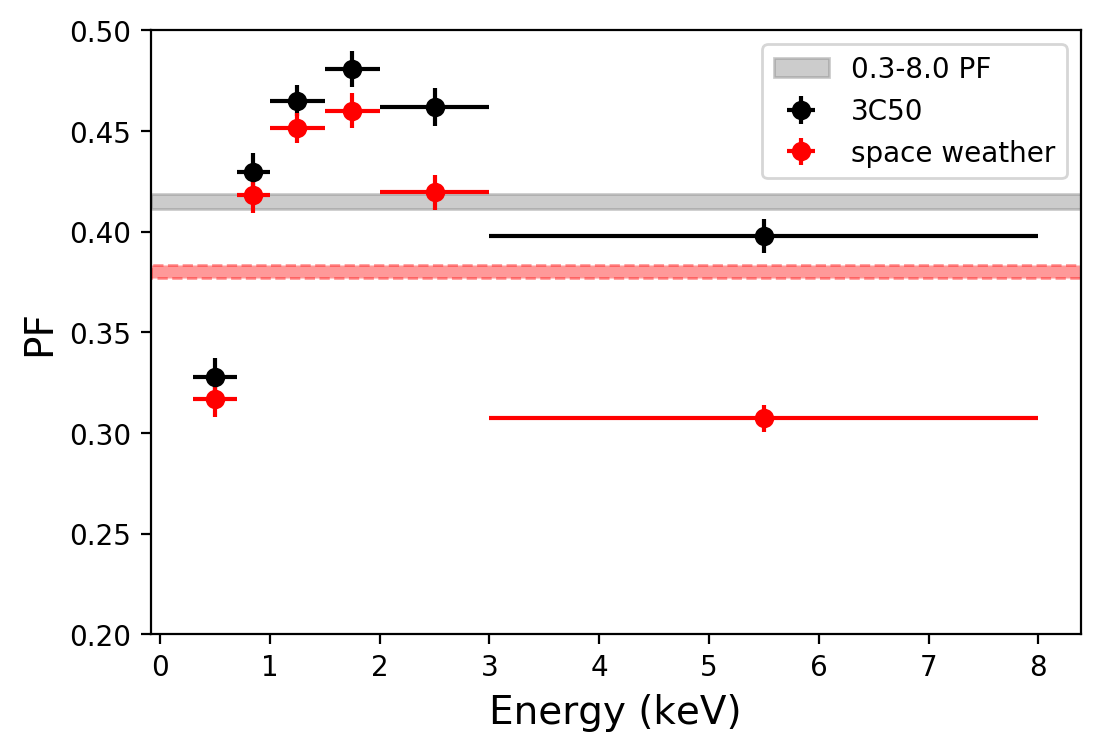}}
\end{minipage}
\begin{minipage}[b]{.49\textwidth}
  \subfloat
    []
    {\label{fig:figD}\includegraphics[width=\textwidth]{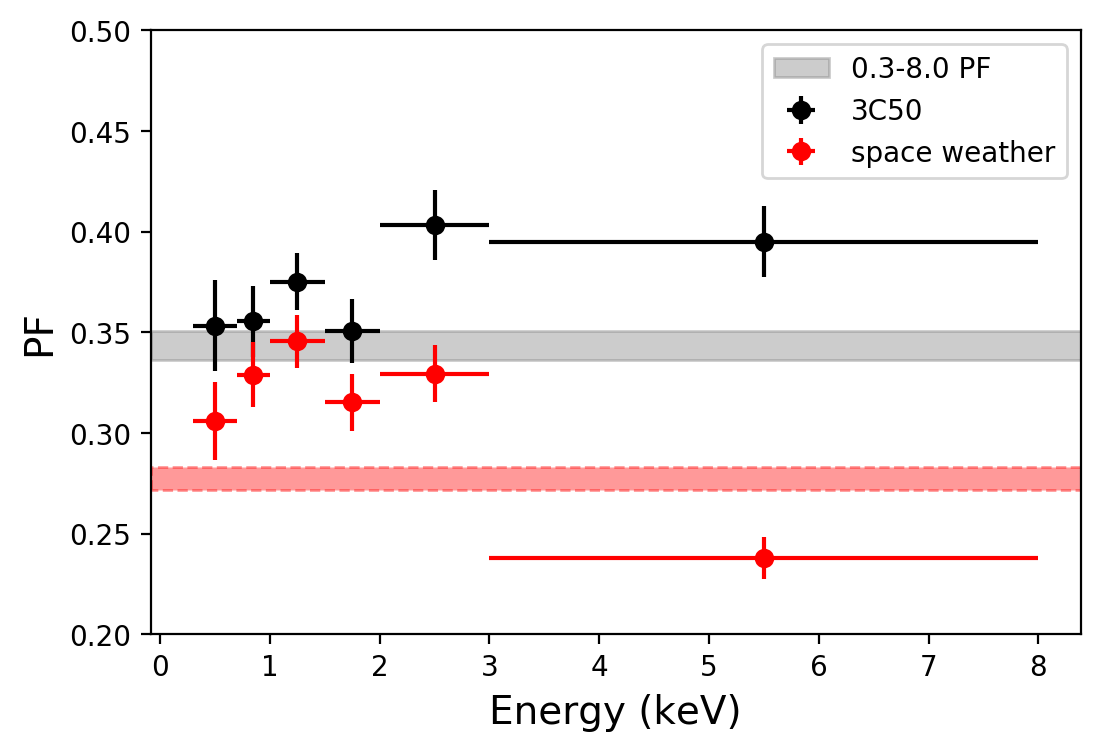}}
\end{minipage}
    \caption{\emph{Left:} temporal properties from \nicer epoch 1 (a) 
    Top: Pulse profiles (with period 68.68s) using 32 phase bins for the total 0.3--8.0 keV count rate as well as the soft (0.3--1.5 keV) and hard (1.5--8.0 keV) bands. Bottom: Pulse phase resolved hardness ratio, calculated using the soft and hard energy bands. Shading shows mean and standard error of HR.
    (c) Energy-resolved rms PF for epoch 1. For the \nicer BG subtraction we used the {\tt 3C50} model for the black points, while for the red points we showed the effect of using the BG determined from space weather. PF and uncertainty for full 0.3--8.0 keV band shown with horizontal stripes.
    \emph{Right:} panels (b) and (d) same as panels (a) and (c), respectively, but for \nicer epoch 2. 
    }
    \label{fig:3panel}
\end{figure*}
\begin{figure}
    \centering
    \includegraphics[width=1.0\columnwidth]{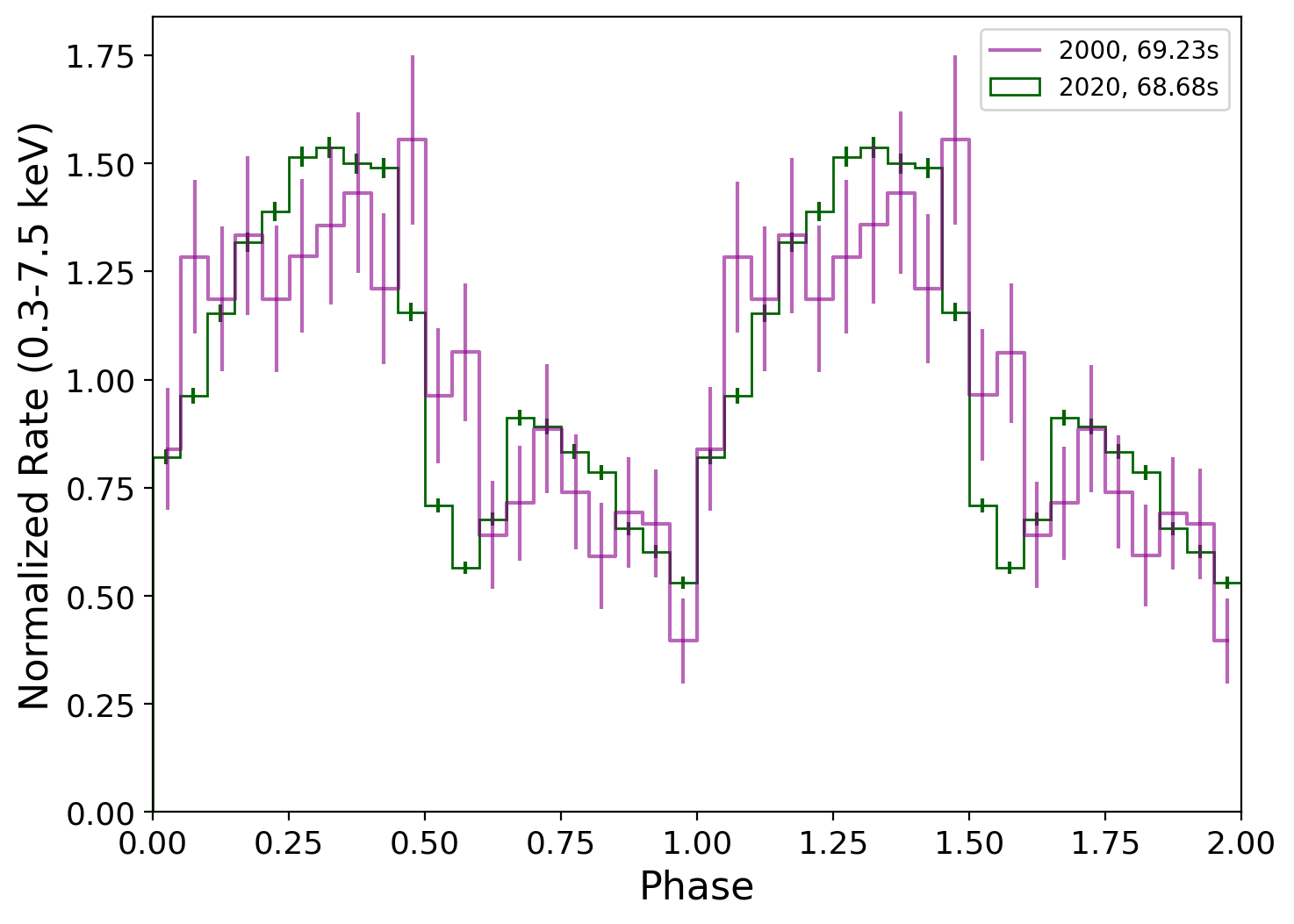}
    \caption{Pulse profiles from 2000 (\xmm, purple) and 2020 (\nicer epoch 1, green), each in the 0.3--7.5 keV band. The 2000 and 2020 data are folded using 69.232 and 68.682 seconds, respectively.  
    }
    \label{fig:20002020}
\end{figure}

We analysed events from epochs 1 and 2 to search for the NS spin period and study the morphology of the pulses.
We first searched for a periodic signal in the barycentric-corrected \nicer events collected within epoch 1 (i.e., MJD 59012.6-59014.1). 
We limited our search within this observing window as multiple \nicer snapshots were obtained, enabling us to phase connect any temporal search. Moreover, the source was at its brightest stage within the same observing window.
We filtered epoch 1 events using standard filtering that resulted in 7~ks of GTI, and with detector energies between 0.3--8.0 keV. 
Pulsations are apparent by-eye with appropriate binning of the events. To measure the most probable period, we used the epoch folding Z-search method implemented through {\tt HENdrics} command-line scripts \citep[][]{2019ApJ...881...39H}.
The epoch folding periodogram is presented in Fig. \ref{fig:zsearch}. 
The gaps in the \nicer data cause aliasing between the correct spin period and the observing window function, resulting in the presence of multiple peaks in the periodogram. 
The most probable solution is found at period P of $\sim$68.682 s, while two symmetric solutions are found at $\sim$67.847 s and $\sim$69.536 s.

To generate a phase-coherent timing model
for the data we used \textsc{Tempo2} \citep{2006MNRAS.369..655H}. We followed the methodology described by \citet{2011ApJS..194...17R}, that has been applied to other slow X-ray pulsars \citep{2019ApJ...879..130R,2019MNRAS.488.5225V,2020MNRAS.494.5350V,2020MNRAS.498.4830J}. 
We subdivided the data into 52 intervals and generated pulse time of arrivals (TOA) by comparing the pulse profile of each interval with a template profile.
We fitted the TOAs to a timing model with one frequency derivative, i.e., $\nu(t)=\nu_{\rm 0}+\dot{\nu}(t-t_{\mathrm 0})$ while setting $t_{\mathrm 0}$ to 59013 MJD.
The best pulse period is consistent with the main peak found with the epoch folding method, while the first derivative is consistent with zero (within 3$\sigma$). 
Nevertheless, the solution has some timing residuals likely caused by torque noise. The results of the coherent timing analysis are summarized in Table \ref{tab:timing}. The temporal solution is in agreement with the one derived from epoch-folding. 
We performed a similar analysis for the \nicer snapshots obtained during epoch 2, between MJD 59023.6 and 59024.5. 
The temporal solution was consistent with the periodicity of 68.57s derived by epoch folding. However, given the smaller total exposure time and lower statistics, no constraints in the period derivative was possible.

\subsection{Pulse Profile Morphology}
\label{sec:pulse_profile}
Folding the epoch 1 events with the best timing solution results in a double-peaked pulse profile with a weaker second peak (see Fig. \ref{fig:3panel}). The two peaks are separated in phase by $\sim$0.4. 
Folding the epoch 2 events with the new best period of 68.57 seconds showed a double-peaked pulse profile (Fig.~\ref{fig:3panel}). In these later measurements, the second peak is more prominent with respect to the primary peak. 
The average background-subtracted count rate decreased from $7.0 \pm 0.4$ to $3.9 \pm 0.2$ c/s between epoch 1 and epoch 2. 

To quantify the strength of the periodic modulation we computed the root-mean-squared (rms) pulsed fraction, which is given by:
\begin{equation}
PF_{\rm RMS}=\frac{\left(\sum_{j=1}^{N} (R_{j} - \bar{R})^2/N\right)^{1/2}}{\bar{R}},
\end{equation}
where N is the number of phase bins, $R_{j}$ is the count rate in the $j$th phase bin, and $\bar{R}$ is the average count rate in all bins \citep[e.g.][]{2018ApJ...863....9W}.
We computed PF for both BG models in order to explore whether the choice of model would affect our results.
For a pulse profile constructed with 32 phase bins, we find PF$_{\rm RMS}$=$0.415 \pm 0.005$ (0.3--8.0 keV), while the energy resolved PF is presented in Fig.~\ref{fig:3panel}.
The overall PF$_{\rm RMS}$ decreases to $0.344 \pm 0.010$ in epoch 2. 
This change in PF, combined with the increase in the intensity of the second peak, causes the decreased variability in the \ero light curve (see Fig. \ref{fig:erosita_lc}), since individual snapshots only sample part of the pulse profile.
For each epoch, the {\tt 3C50} yields a higher PF than the space weather-based BG model. 
Using either BG model for epoch 1, PF appears to have a maximum value around 2 keV followed by a decrease in PF with energy.
A decrease in PF at higher energies (3--8 keV) is also found in epoch 2 according to the space weather model. However, with the preferred {\tt 3C50}, this behavior is not replicated.
The energy-resolved PF behavior is, of course, dependent on the definition of PF used \citep[see examples by][]{2018MNRAS.479L...1Y}. For example, if we define PF as $\rm{PF}=(\rm{R}_{Max}-\rm{R}_{Min})/(\rm{R}_{Max}+\rm{R}_{Min})$, the {\tt 3C50} model in epoch 1 does not maintain the behavior of a decrease in PF after 2 keV. In fact, by defining PF based on minimum and maximum rate only, we find almost stable PFs between 2--8.0 keV, and consistent values between the two epochs.

To complement our study of the pulse morphology we also searched for evidence of spectral evolution within pulse phase. To do that we used the hardness ratio (HR), defined as $\rm{HR}=(\rm{R}_{i+1}-\rm{R}_{i})/(\rm{R}_{i+1}+\rm{R}_{i})$, where $\rm{R}_{i}$ is the count rate in a specific energy band $i$. We computed HR using the 0.3--1.5 keV and 1.5--8.0 keV energy bands and 32 phase bins (see Fig. \ref{fig:3panel}). 
We note that the hardness ratio increased from epoch 1 to epoch 2; its behavior as a function of phase is also included in Fig.~\ref{fig:3panel}. Both epochs show dips in HR at the count minima following the primary peak. Only the first epoch clearly displays this behavior following the secondary peak as well.

The pulse profile constructed with \nicer data from epoch 1 is similar to \xmm profile from quiescence in observations from 2000 \citep{2003A&A...406..471H}. In Fig.~\ref{fig:20002020}, the double-peaked profiles are mostly consistent, despite the system's higher luminosity during the peak of the outburst by a factor of $\sim$130. Furthermore, during the decay of the June 2020 outburst, when the luminosity had only decreased by $\sim$35\%, the pulse profile morphology had clearly changed (Fig.~\ref{fig:3panel}), with a strengthened second peak, making the morphology no longer consistent with the previous two profiles. This behavior is further discussed in Sect. \ref{sec:supercritical}.

\begin{figure*}
    \centering
    \includegraphics[width=0.45\textwidth]{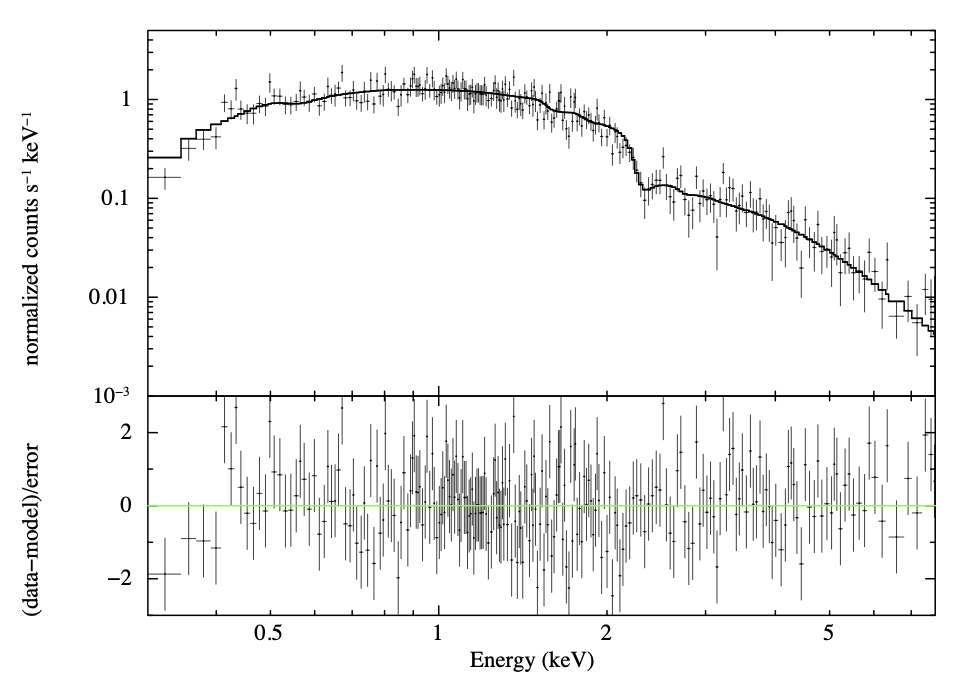}
    \includegraphics[width=0.45\textwidth]{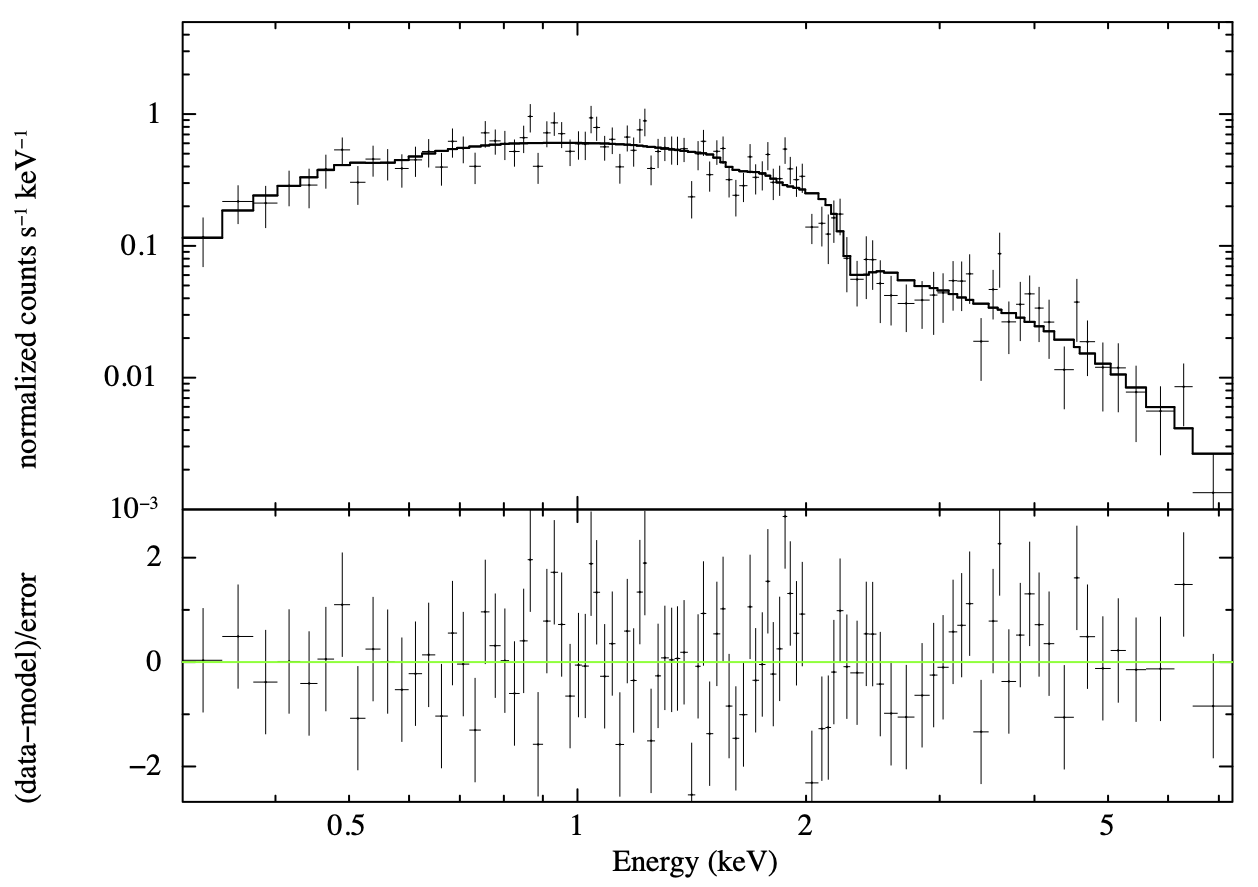}
    \includegraphics[width=0.45\textwidth]{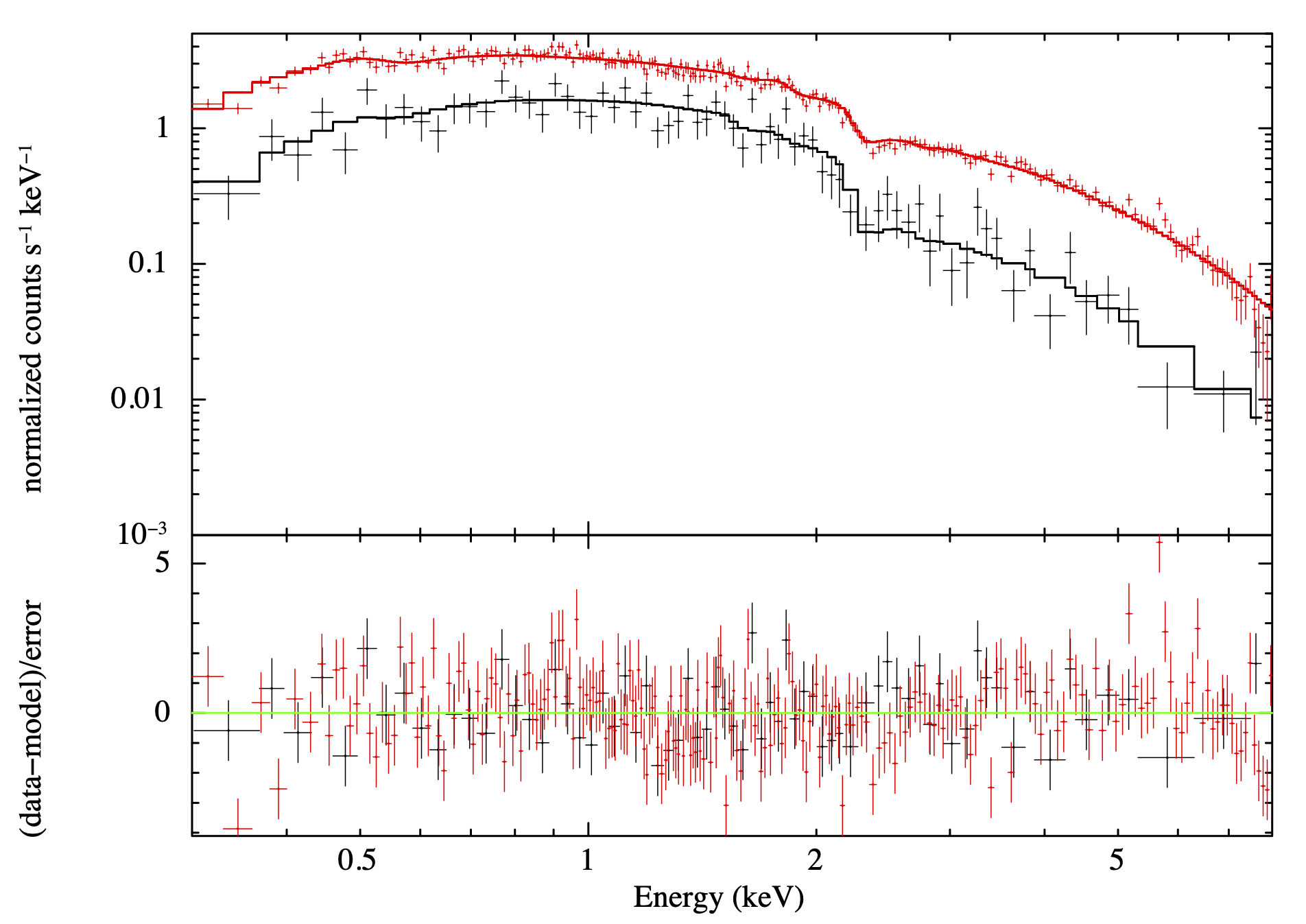}
    \includegraphics[width=0.45\textwidth]{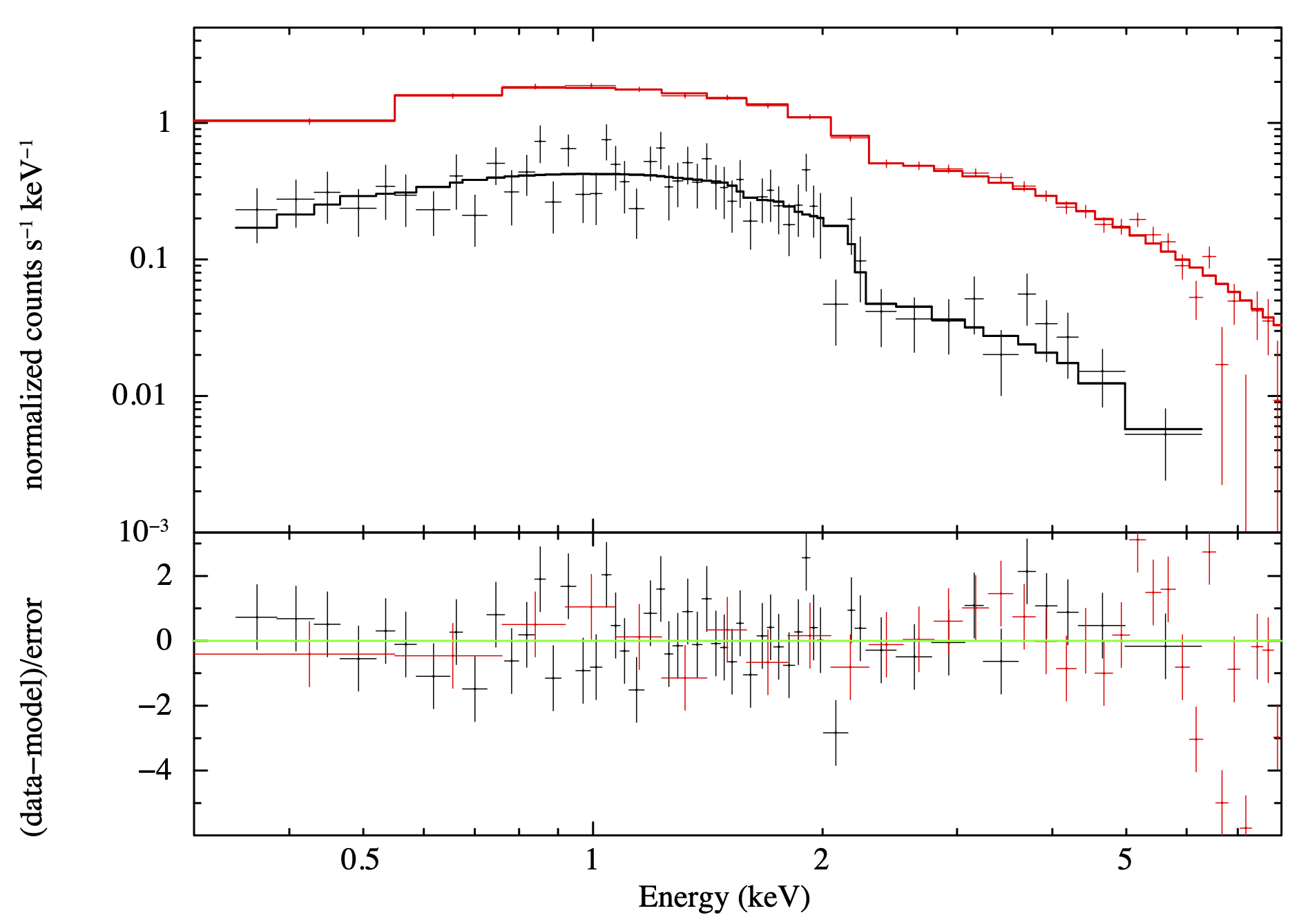}
    \caption{Top: \ero spectra during its first epoch (left) and second epoch (right). Bottom:  \ero spectra obtained during the epoch of the first and second \nicer observations, with corresponding \nicer spectra in red. Spectra were fit independently.}
    \label{fig:spec2}
\end{figure*}

\subsection{Spectral properties}
\label{sec:spectra}
In BeXRBs, hard X-rays originate from the accretion column, and can be fitted by a phenomenological power law-like shape with an exponential high energy cutoff above (or around) 10 keV \citep[e.g.,][]{2013A&A...551A...6M,2014MNRAS.444.3571S,2013A&A...558A..74V,2016MNRAS.461.1875V}. In many cases BeXRB spectra show residuals at soft energies. These residuals are often referred to as a ``soft-excess'' whose physical origin is attributed to a combination of mechanisms like emission from the accretion disk, emission from the NS surface, or hot plasma around the magnetosphere \citep{2004ApJ...614..881H}.

Motivated by the changes in the timing properties, we separately fitted the spectra from the two epochs to investigate potential spectral changes. Moreover, to investigate instrumental differences, we extracted subsets of \ero data taken contemporaneously with the \nicer observations.
Spectral analysis was performed using {\tt XSPEC} v12.10.1f \citep{1996ASPC..101...17A}.

Given the lack of broadband spectra we used a power law model to fit the continuum. 
To account for the photoelectric absorption by the interstellar gas we used {\tt tbabs} in {\tt xspec} with solar abundances set according to  \citet{2000ApJ...542..914W} and atomic cross sections from \citet{1996ApJ...465..487V}. 
Column density was left as a free parameter with solar abundances. To separately fit Galactic and intrinsic absorption -- a typical procedure for BeXRBs in the MCs -- we tested using two absorption components \citep[e.g.,][]{2013A&A...558A..74V,2016MNRAS.461.1875V,2018MNRAS.475..220V}. 
The free component is meant to model absorption within the LMC and near the source, so elemental abundances are fixed at 0.49 solar \citep{2002A&A...396...53R}. Because this component is consistent with zero and not constrained, we used only the free absorption component in the reported fit. 
We found that the derived column density from the spectra is in agreement with the Galactic value of $5.65 \times10^{20}$ cm$^{-2}$ \citep{1990ARA&A..28..215D}, implying that no significant absorption is present around the source or in its vicinity. While finding no additional absorption from LMC material is typical for many northern LMC sources, particularly in the area around the supergiant shell $[$KDS99$]$~SGS 11 \citep{1999AJ....118.2797K,2016A&A...585A.162M}, the lack of absorption intrinsic to the source or to its immediate vicinity is quite uncommon for BeXRBs.

For the spectral fit of \ero data we used C-statistics \citep{1979ApJ...228..939C},
while the \nicer spectra were fitted independently by minimizing the {\tt pgstat} statistic\footnote{{\tt pgstat} is suited for Poisson distributed data with Gaussian distributed BG; see \url{https://heasarc.gsfc.nasa.gov/xanadu/xspec/manual/XSappendixStatistics.html}} in {\tt xspec}.
Through spectral fitting, we found that the \nicer BG estimates based on the ``space weather'' method yield lower column density values for the absorption component and a higher flux ($\sim$25--30\%) than the contemporaneous \ero spectra.
Moreover, the \nicer model-to-data residual plots showed a high energy excess (>5keV) that significantly differed from the \ero ones. Thus, based on these characteristics, we conclude that the {\tt 3C50} BG model better reproduces the actual BG during the \nicer observations of \lxp. As such, the {\tt 3C50} BG model is used for Fig.~\ref{fig:3panel}, although energy-resolved PFs are plotted using both BG models to illustrate the effect of model choice.
The spectra with the best fit models are shown in Fig. \ref{fig:spec2}.
The derived fluxes were converted to unabsorbed luminosities by correcting for absorption effects and for the distance of 50 kpc to the LMC. The results are presented in Table \ref{tab:spectral}.  
Both instruments show that \lxp is softer when brighter, since the photon index of the spectra decreases from the first to the second epoch.

The power-law slope is within the range of 0.6-1.4 that is typically observed in BeXRB outbursts  \citep{2008A&A...489..327H,2016A&A...586A..81H}. Moreover, we investigated the spectra for the presence of an Fe K$\alpha$ emission line that is typically observed in accreting pulsars. There is some structure around 6.4 keV in the residual \nicer spectrum during epoch 1. We used a Gaussian component to model this feature, but we found no statistical evidence of an emission line. By fixing the line central energy and width at 6.4 keV and 0.1 keV, we established an upper limit of $\sim$120 eV for the equivalent width of the line. This value is not surprising as BeXRB pulsars in the LMC can typically show Fe K$\alpha$ lines with equivalent widths below 100 eV \citep[e.g.][]{2013A&A...558A..74V,2014A&A...567A.129V}.  

As noted in Section \ref{sec:pulse_profile}, the phase-resolved HR plots for each \nicer epoch suggest a dip in hardness at a phase of $\sim$0.6. We extracted the \nicer spectrum within 0.1 of that phase to test if there is a significant difference in the spectral parameters. 
The resulting spectrum had $\sim$3000 net counts (i.e., from source and background), and was fitted with the same absorbed power law model.
The fit gave $0.97^{+0.10}_{-0.03}$ for $\Gamma$, which is consistent with the overall \nicer fit. On the other hand, the $N_H$ of $0.2^{+1.7}_{-0.1}$ $10^{20}$cm$^{-2}$ is lower than the overall best values.
A phase-dependent absorption could be indicative of absorption due to material located in the NS magnetosphere. However, we cannot exclude that this spectral softening could be related to a soft excess that is invariant with spin phase. A non-pulsating thermal component with spectral peak below 1 keV could result in a spectral softening. Nevertheless, as we indicated above, we cannot constrain such component with the given spectral statistics.
\begin{table*}
\caption{Best-fit parameters of spectral models}\label{tab:spectral}
\begin{threeparttable}[b]
\begin{tabular*}{\textwidth}[t]{p{0.09\textwidth}p{0.1\textwidth}p{0.10\textwidth}p{0.10\textwidth}p{0.10\textwidth}p{0.10\textwidth}p{0.10\textwidth}p{0.17\textwidth}}
\hline
\hline\noalign{\smallskip} 
 & \multicolumn{3}{l}{Epoch 1} & \multicolumn{3}{l}{Epoch 2}  &  \\
Parameter  & eROSITA & eROSITA & NICER & eROSITA & eROSITA & NICER  &  Units \\

\hline\noalign{\smallskip}  
Dates  $^{(a)}$ & 5-16.5 & 12.7-14&  12.6-14.1 & 19.4-29& 22-25& 13.6-24.5 &  days \\
\hline\hline\noalign{\smallskip}  
$N_{\rm{H}}$ & 4.0$^{+1.6}_{-1.6}$ &  $4.1^{+2.8}_{-2.3}$ & $4.26^{+0.54}_{-0.50}$ & $3.8^{+3.4}_{-2.6}$ &  $<0.003$$^{(b)}$ & $7.4^{+1.2}_{-1.2}$ &$10^{20}$cm$^{-2}$\\\noalign{\smallskip} 
$\Gamma$  & 0.95$^{+0.07}_{-0.07}$ & 0.98$^{+0.11}_{-0.12}$ &  0.97$^{+0.02}_{-0.02}$ & 0.89$^{+0.16}_{-0.12}$ & 0.64$^{+0.40}_{-0.08}$ & 0.89$^{+0.04}_{-0.04}$&\\\noalign{\smallskip}  
$F_\mathrm{{X}}$  $^{(c)}$ & $2.63^{+0.16}_{-0.15}$&$2.67^{+0.26}_{-0.23}$  & $2.72^{+0.04}_{-0.05}$ &$1.68^{+0.20}_{-0.19}$ &$1.71^{+0.29}_{-0.48}$  &$1.76^{+0.05}_{-0.06}$ & \oergcm{-11} \\\noalign{\smallskip} 
\hline\noalign{\smallskip}  
$\mathrm{Fit}_{\rm stat}/dof$ $^{(d)}$ & 459.49/546 & 332.96/352 & 817.81/769 &367.79/365 & 202.94/216 & 779.04/769 & \\\noalign{\smallskip}  
$L_\mathrm{{X}}$ $^{(e)}$ & $7.86^{+0.48}_{-0.44}$& $7.98^{+0.77}_{-0.68}$ & $8.13^{+0.13}_{-0.14}$ & $5.02^{+0.59}_{-0.56}$ & $5.1^{+0.9}_{-1.4}$ &$5.26^{+0.16}_{-0.17}$ & \oergs{36} \\\noalign{\smallskip}
\hline\noalign{\smallskip}  
\end{tabular*}
\tnote{(a)} Interval used for spectral extraction, dates in MJD-59000. 
\tnote{(b)} Not well-constrained by fit
\tnote{(c)} X-ray flux in the 0.3--8.0 energy band
\tnote{(d)} \ero and \nicer spectra were fitted with C-stat and \texttt{pgstat} statistics respectively.
\tnote{(e)} Absorption corrected X-ray luminosity in the 0.3--8.0 energy band, estimated for a distance of 50 kpc.
\end{threeparttable}
\end{table*}

\subsection{Long-Term X-ray LC}
\label{sec:swift_data}

During the decay of the 2020 outburst, \lxp was also monitored by the
Neil Gehrels \swift\ Observatory \citep[\swift,][]{2004ApJ...611.1005G} X-ray Telescope \citep[XRT, ][]{2005SSRv..120..165B} between MJD 59031.3--59059.
To investigate the long-term variability of the system and compare it with the available optical data (see next section) we searched for other available archival X-ray data.
The location of the source has been observed a few times by \swift/XRT. 
To calculate average count rates for all \swift/XRT detections, we analysed available data through the \swift science data centre following \citet{2007A&A...469..379E,2009MNRAS.397.1177E}. 
During these observations, \swift also obtained data with the Ultraviolet/Optical Telescope \citep[UVOT,][]{2005SSRv..120...95R}, 
in several of the six broad-band UVOT filters (\textit{v}, \textit{b}, \textit{u}, \textit{uvw1}, \textit{uvm2}, \textit{uvw2}) described in \citet{2008MNRAS.383..627P,2011AIPC.1358..373B}. 
The UVOT images were analysed using the \texttt{uvotsource} tool included in {\sc heasoft} using standard procedures. Given the lack of ground-based monitoring of the LMC during the outburst, these are the only available optical data during this period.

Moreover, the LMC field was observed once by \xmm (obsid: 0841320301, PI: P. Maggi) in February 2020 ($\sim$MJD 58893.5), a few months before the 2020 outburst. \xmm data were analysed with a standard pipeline developed for the MCs \citep{2013A&A...558A...3S,2019A&A...622A..29M}. \lxp was detected by EPIC-pn with a count rate of 0.048$\pm$0.02 c/s in the 0.2--12\,keV band. The \xmm observation was affected by high background, which resulted in only a few hundred source counts within good time intervals, making no detailed timing and spectral analysis possible.

\begin{figure*}
    \centering
    \includegraphics[width=1.0\textwidth]{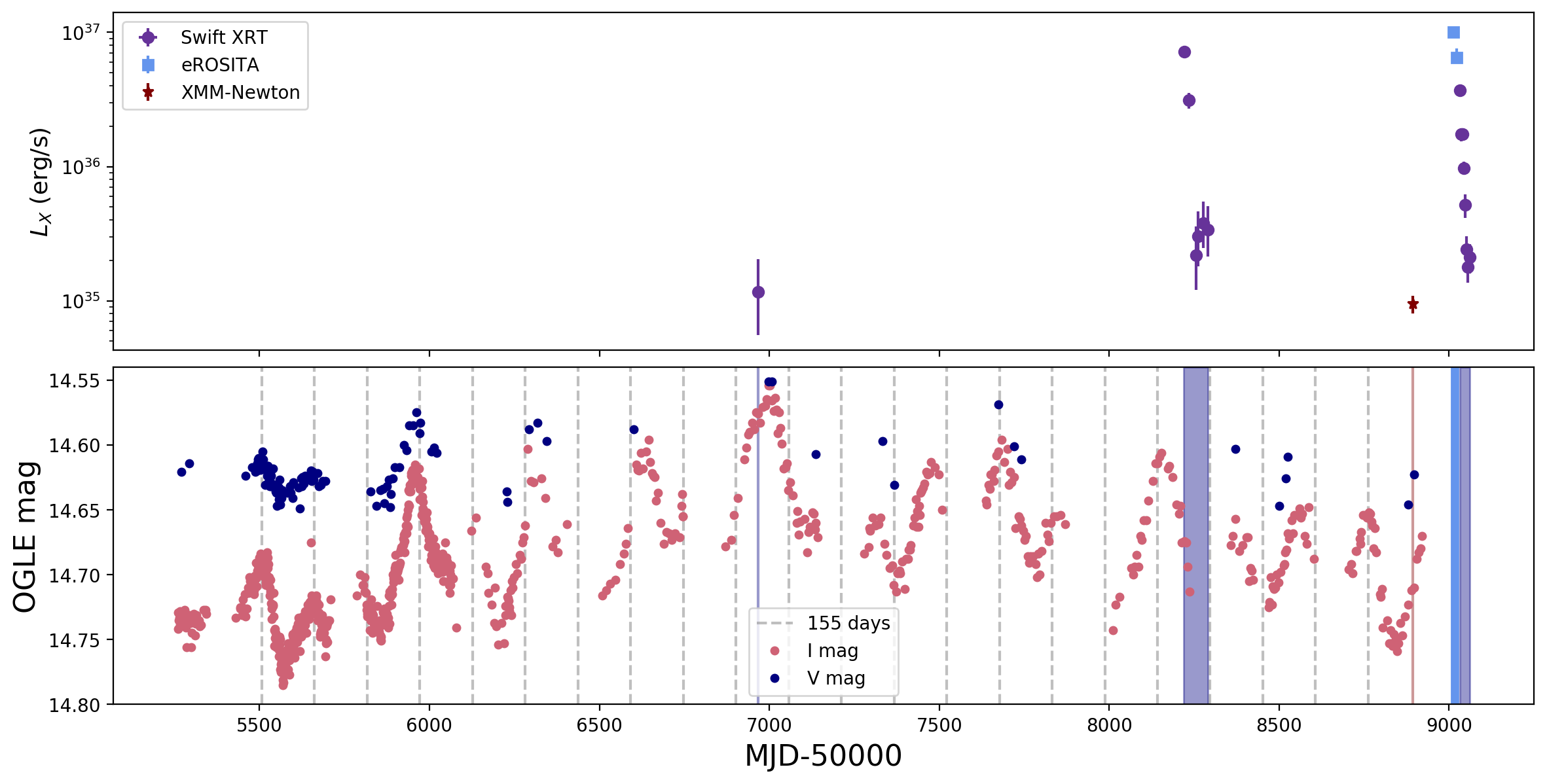}
    \caption{Top: Long-term X-ray luminosity (0.3--10.0 keV) from \swift, \ero, and \xmm using one blue square each for the epochs before and after the gap in the data shown in Fig.~\ref{fig:erosita_lc}. Bottom: OGLE I and V band data, spanning from March 2010 to March 2020. Shading shows ranges when X-ray data, shown above, was also taken. Grey dashed lines are spaced by 155 days from the first flare center. It is clear that several optical peaks do not align with the derived period from the LS periodogram.}
    \label{fig:fulllcs}
\end{figure*}

To convert count rates from all other instruments to unabsorbed $L_{X}$ in the 0.3--10.0 keV band we adopted the best fit models from the two \ero epochs (see Table \ref{tab:spectral}). In particular, we used the different spectral parameters ($N_H$ and $\Gamma$) from epochs 1 and 2 to infer upper and lower bounds when converting from count rates to flux.
The historic X-ray light curve of \lxp is shown in Fig.~\ref{fig:fulllcs}. In the same figure, we compare the long-term X-ray behavior of \lxp to the OGLE I and V band light curves, which span March 2010--March 2020.
In Sect. \ref{sec:ogle} we present detailed analyses of optical data and discuss the relationship between the X-ray and optical behavior. 

\begin{figure*}
    \centering
\begin{minipage}[b]{.46\textwidth}
  \subfloat
    []
    {\label{fig:LSfigA}\includegraphics[width=\textwidth,trim={0 0 0 0},clip]{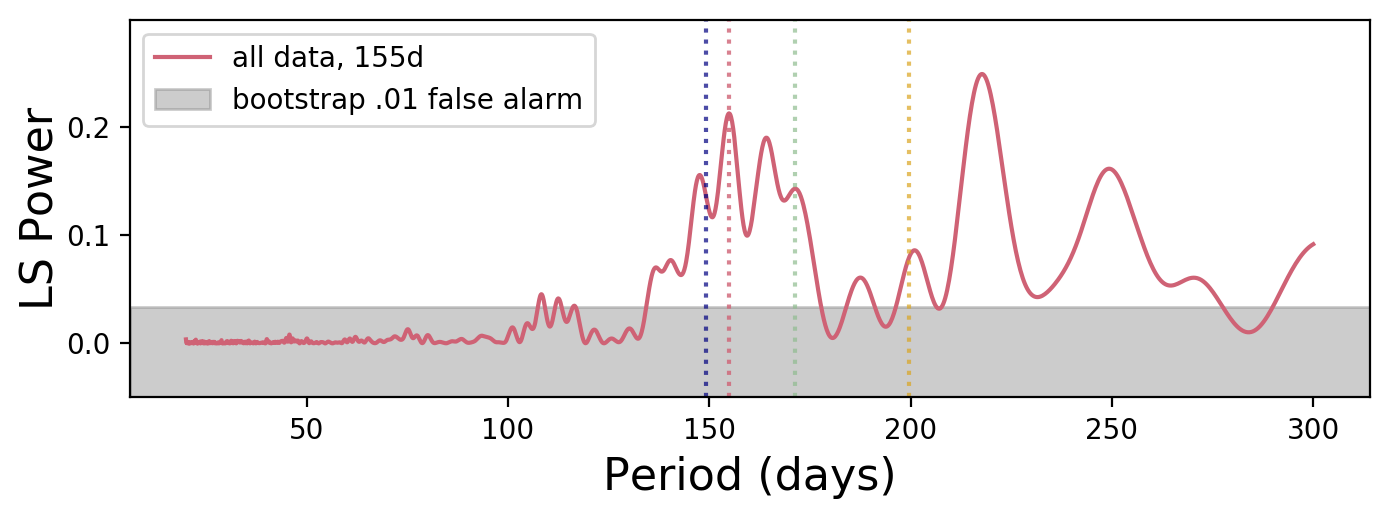}}
\end{minipage}
\begin{minipage}[b]{.46\textwidth}
  \subfloat
    []
    {\label{fig:LSfigB}\includegraphics[width=\textwidth,trim={0 0 0 0},clip]{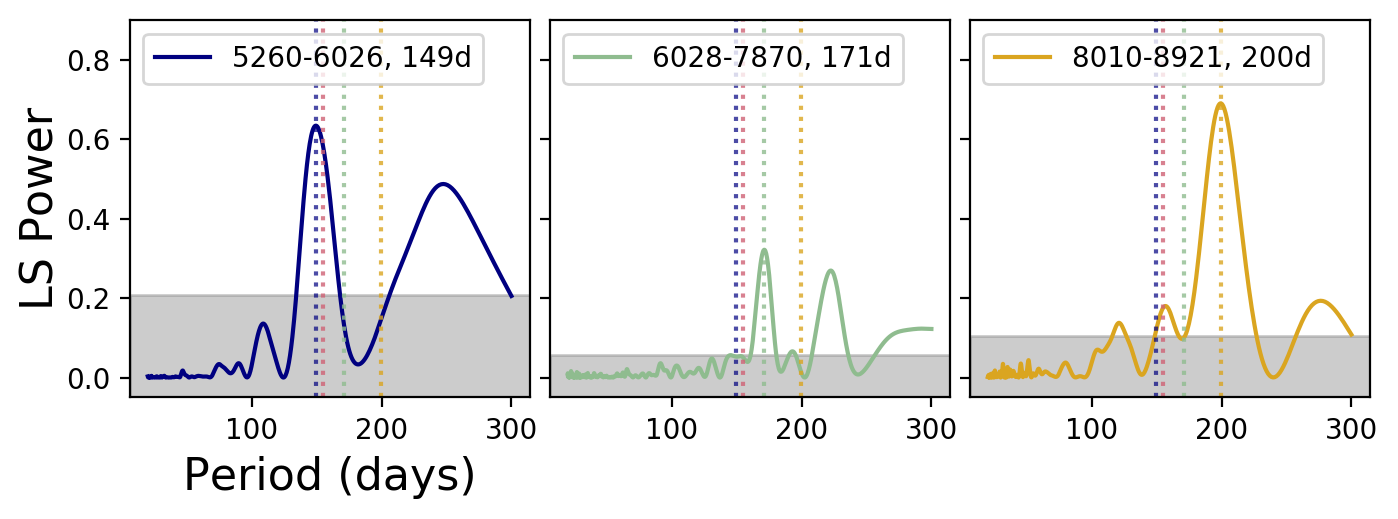}}
\end{minipage}

\begin{minipage}[b]{.95\textwidth}
  \subfloat
    []
    {\label{fig:LSfigC}\includegraphics[width=\textwidth,trim={0 0 0 0},clip]{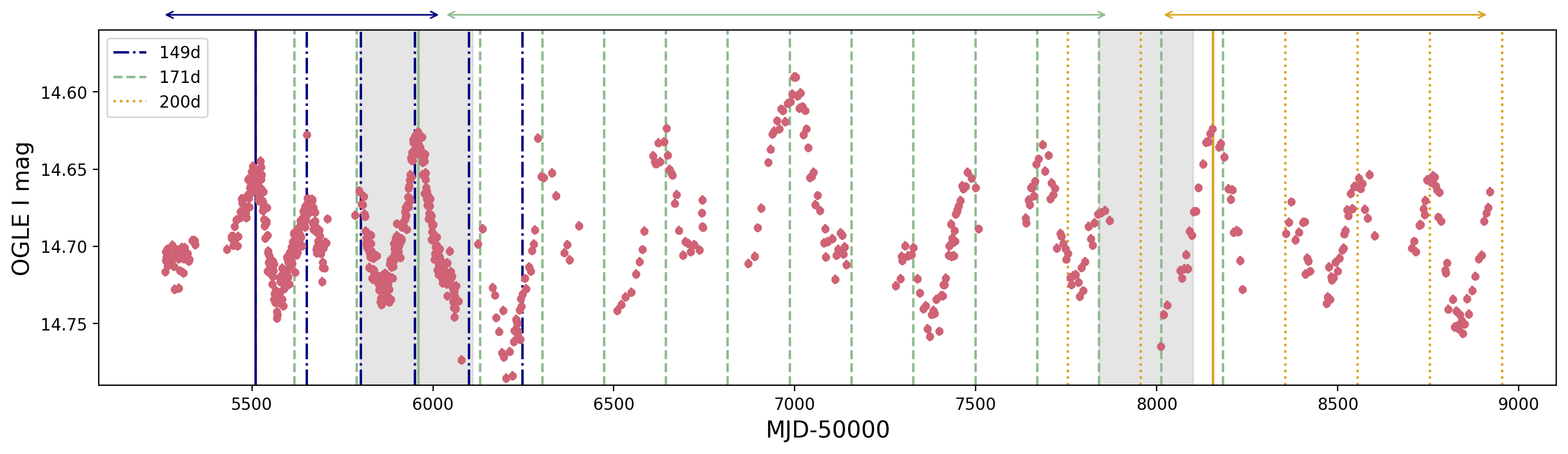}}
\end{minipage}
\caption{(a) LS periodogram derived from the I band LC. (b) Three panels show the LS periodograms derived from three epochs. It is clear that the stronger periodic signal varies between complete LC and individual epochs. For comparison we have plotted the same vertical lines in each sub-figure. (c) Detrended I band LC. Vertical lines indicate the best estimated period from the LS periodogram for three different epochs. Period predictions begin from flare centers marked with solid lines. ``Transition'' regions are shaded in gray, with arrows denoting epoch boundaries. Period predictions are extended beyond each epoch to indicate the presence of sharp rather than gradual transitions.}
    \label{fig:OGLE_ALL}
\end{figure*}

\section{OGLE: 10 years of optical monitoring}
\label{sec:ogle}
The Optical Gravitational Lensing Experiment (OGLE) started operations in 1992 \citep{1992AcA....42..253U, 2008AcA....58...69U}. The LMC is monitored by OGLE, with observations made with the 1.3\,m Warsaw telescope at Las Campanas Observatory. Images are taken in the V and I bands. The data reduction is described in \citet{2008AcA....58...69U}. The OGLE database provides access to optical light curves of most identified X-ray binaries in the MCs\footnote{OGLE XROM portal: \url{http://ogle.astrouw.edu.pl/ogle4/xrom/xrom.html}}, so OGLE data have been crucial in the study of X-ray binaries in the nearby galaxies  \citep[e.g.,][]{2013A&A...554A...1M,2017A&A...598A..69H,2018MNRAS.475.3253V,2020MNRAS.499L..41K,2020MNRAS.499.2007V}

The OGLE optical counterpart of \lxp was identified as a $\sim$14.5 mag star with coordinates $\alpha{=}$05:29:47.84 $\delta{=}-65$:56:43.7 \citep{2020ATel13828....1H}. The LMC field was monitored by OGLE between March 2010 and March 2020, within which the \lxp optical counterpart shows strong outburst activity on top of a gradual long-term trend of rising then falling overall optical luminosity \citep{2020ATel13828....1H}. In this section we will report on the optical properties and compare the optical and X-ray variability.

\begin{figure}
    \centering
    \includegraphics[width=1.0\columnwidth]{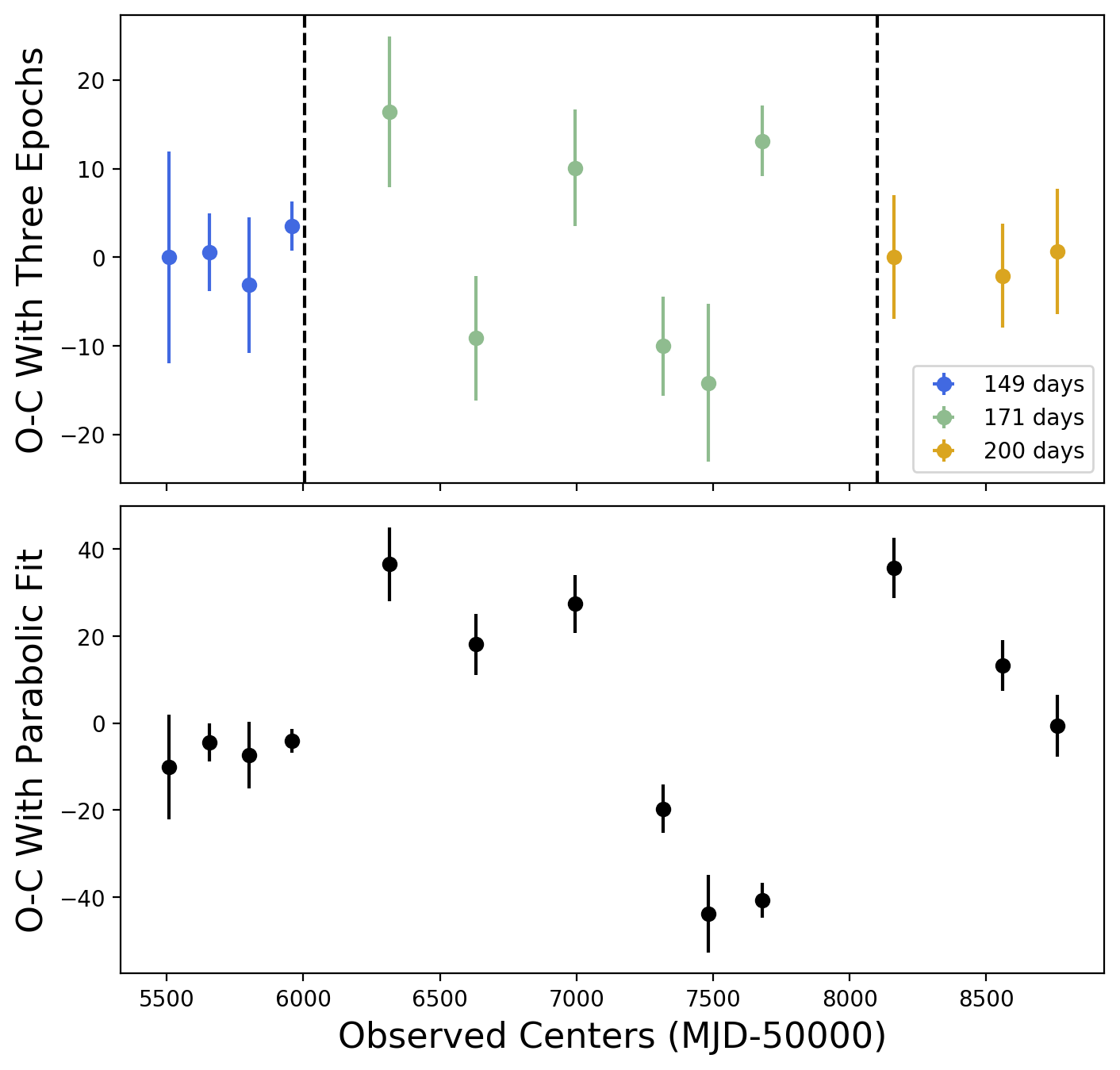}
    \caption{Top: Residual observed - calculated (O-C) plots with calculated centers using 149, 171, and 200 day periods. These period values make better predictions than the linearly changing period (bottom). We do not see a trend of error within each epoch. Increasing O-C over time would support a linearly increasing period.}
    \label{fig:bothoc}
\end{figure}

\begin{figure}
    \centering
    \includegraphics[width=1.0\columnwidth]{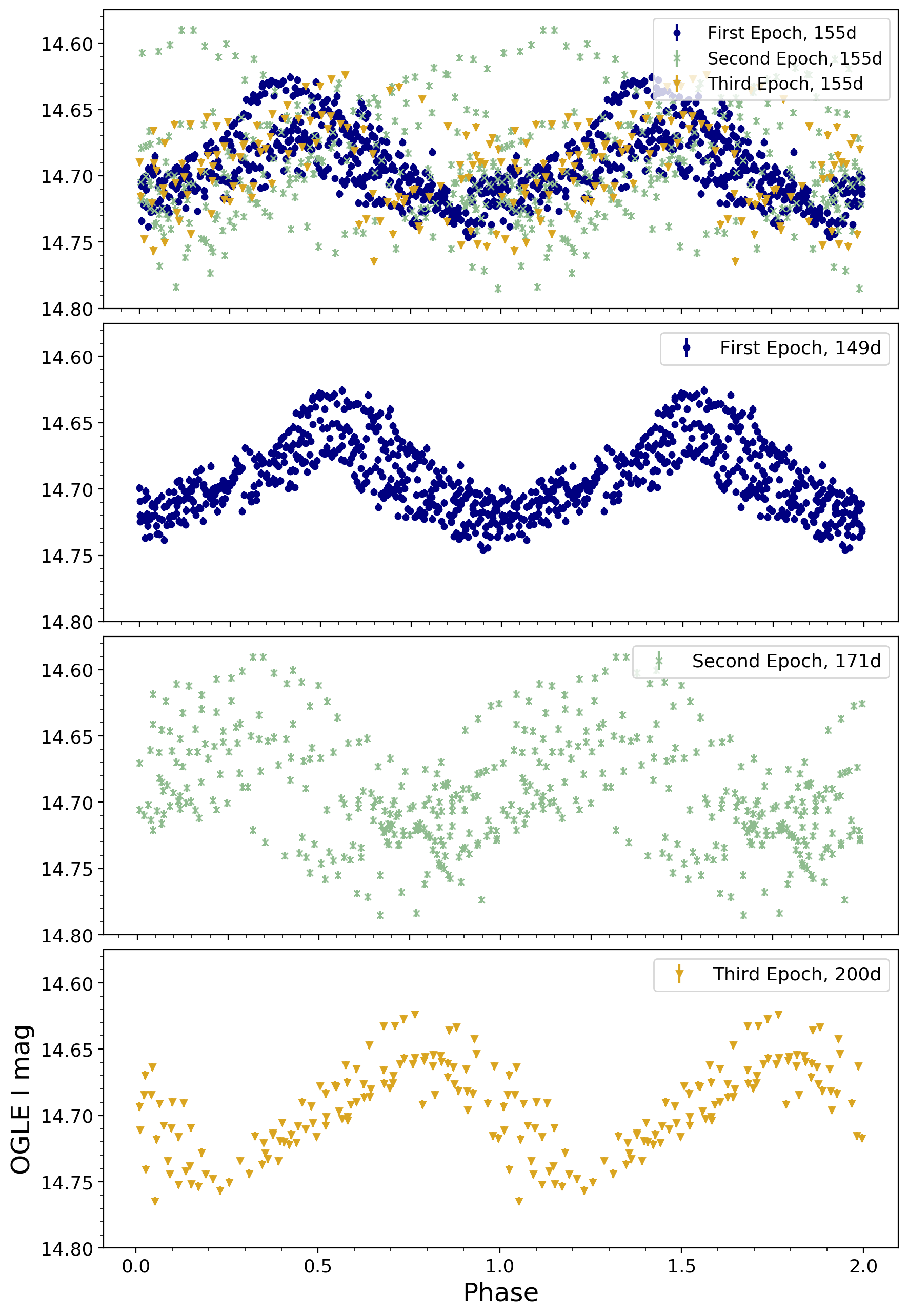}
    \caption{Top: all OGLE detrended I band data phase-folded using the LS peak of 155 days. The bottom three plots show the data within each epoch phase-folded using their respective best periods (149, 171, and 200 days). Note the change in flare shape, especially from the first to the third epoch.}
    \label{fig:4fold}
\end{figure}

\subsection{An Evolving Optical Periodicity}
The full OGLE light curve is shown in Fig.~\ref{fig:fulllcs}. We used the I band data for periodicity search, but incorporated the sparser V band data to investigate V-I colour changes over time. A long-term trend that spans at least the entire range of data (3661 days) is apparent. We expected the repeated flaring to reveal the orbital period, which in turn is helpful in predicting periastron passages. 
We detrended the I band data by subtracting a smoothed light curve derived by applying a Savitzky-Golay filter \citep{1964AnaCh..36.1627S} with a window length larger than the period to make sure peak shape remains unaffected \citep[see similar applications to OGLE data,][]{2012MNRAS.423.3663B,2014MNRAS.444.3571S,2016MNRAS.461.1875V,2017MNRAS.470.1971V}. We then computed a Lomb-Scargle periodogram\footnote{We run the LS alogrithm using both flux and magnitudes, with the methods yielding identical results.} \citep[LS,][]{1982ApJ...263..835S}.
The dominant reasonable peak in the LS periodogram for these data is at 155 days, while a secondary peak appears at 164 days. 
The optical LC has an almost sinusoidal variation, with multiple peaks of various amplitudes. However, individual peaks are not evenly spaced, and appear to be quasi-periodic, which is consistent with the periodogram shape.  
To better visualize this quasi-periodic behavior, we space vertical lines by either 155 or 164 days starting from the first optical peak (at 55507 MJD). We then see that no single period will match the optical maxima. In Fig.~\ref{fig:fulllcs}, it is clear that 155 days is greater than the peaks spacing in the first few optical cycles, but underestimates the spacing between the rest of the cycles. 
Thus, accurate predictions require a changing period over time. This quasi-periodic behavior must be representative of more than the system orbital period, indicating that effects such as disc truncation and precession are at play \citep[e.g.,][]{2011MNRAS.416.2827M,2012ASPC..464..177O}.

To further investigate the evolution of the optical periodicity we modeled individual optical peaks with simple analytical models that allow us to determine the epoch of the optical maximum of each peak.
Moreover, given that there were no OGLE data during the June 2020 X-ray outburst, this modeling enables extrapolation of the OGLE LC and comparison of the occurrence of an X-ray outburst with the optical phase. 
We found center values and uncertainties for 13 optical peaks in the light curve (see Appendix \ref{sec:flarefit}). We then fit a quadratic function to each peak number vs. time. The resulting linearly increasing period did not prove to be the best model. Instead, we can describe three epochs with periods determined using separate LS periodograms: 149, 171, and 200 days, respectively. These same values can be determined with a least-squares method using the modeled and predicted centers within each epoch. The residuals for each model are shown in Fig.~\ref{fig:bothoc}. 
The phase-folded data from each epoch is displayed in Fig.~\ref{fig:4fold}, showing the failure of a single period value and the improvement of phase-folding upon epoch separation.

Inspection of the light curve is necessary to confirm any LS peaks. We found that folding the data with the peaks between 200 and 300 days from Fig.~\ref{fig:OGLE_ALL} resulted in spurious profiles. The peaks are likely the result of the evolving period, differing flare widths, or nonuniform sampling.

\subsection{V-I Colour}
\begin{figure}
    \centering
    \includegraphics[width=1.0\columnwidth]{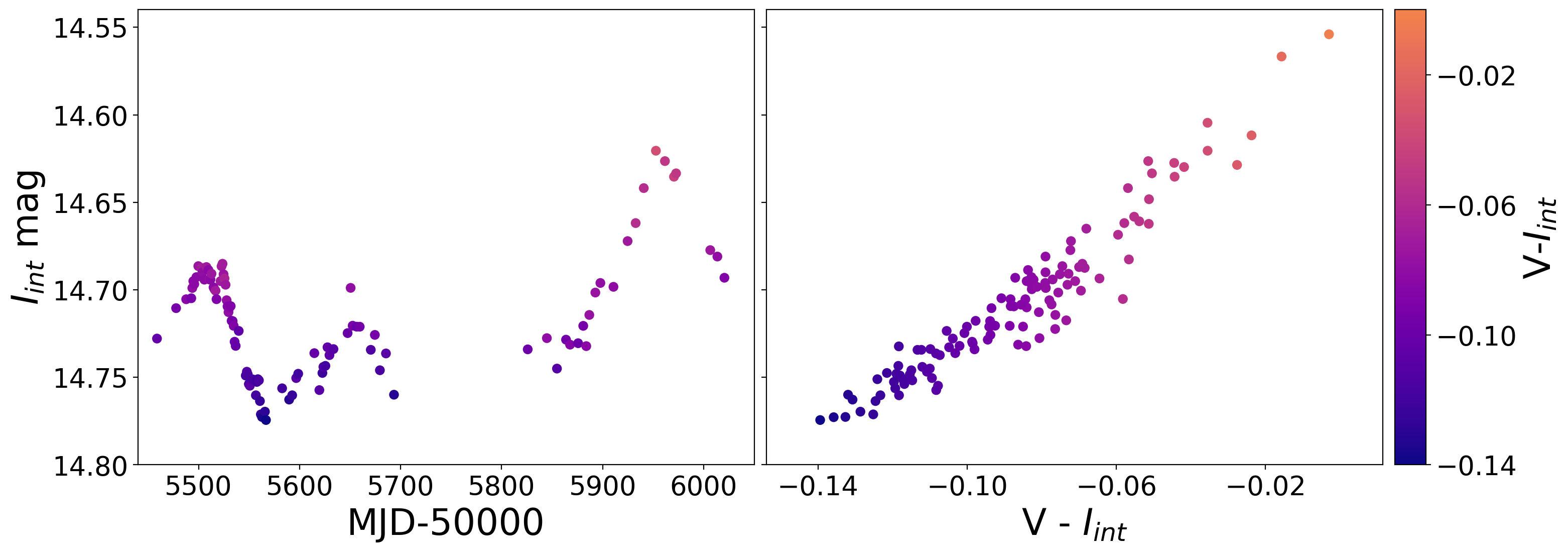}
    \caption{Left: Portion of interpolated I vs. MJD-50000 curve. Right: All interpolated I band plotted against V - $I_\mathrm{{int}}$. Colour shows V - $I_{int}$ in both.}
    \label{fig:colormag}
\end{figure}

In order to obtain V-I values, we linearly interpolated the I band data for an estimate of values at the times of V band observations. Interpolation in I band is preferable since the data is denser than in V band. 
The colour-magnitude relationship is shown in Fig.~\ref{fig:colormag}. 
V-I$\mathrm{_{int}}$ variation is correlated with the V and I light curve changes themselves. This behavior is indicative of a ``redder when brighter" relationship. The brightness increase is driven by disc growth, which provides an infrared excess compared to the stellar photosphere 
\citep[ e.g.,][]{2017A&A...598A..69H,2014A&A...567A.129V,2012MNRAS.424..282C} in both the short and longer-term trend. At the same time, it speaks to the different ranges of variability for the two bands (see Fig.~\ref{fig:fulllcs}). 

\subsection{Optical Behavior During 2020 X-ray Outbursts}
\begin{figure}
    \centering
    \includegraphics[width=1.0\columnwidth]{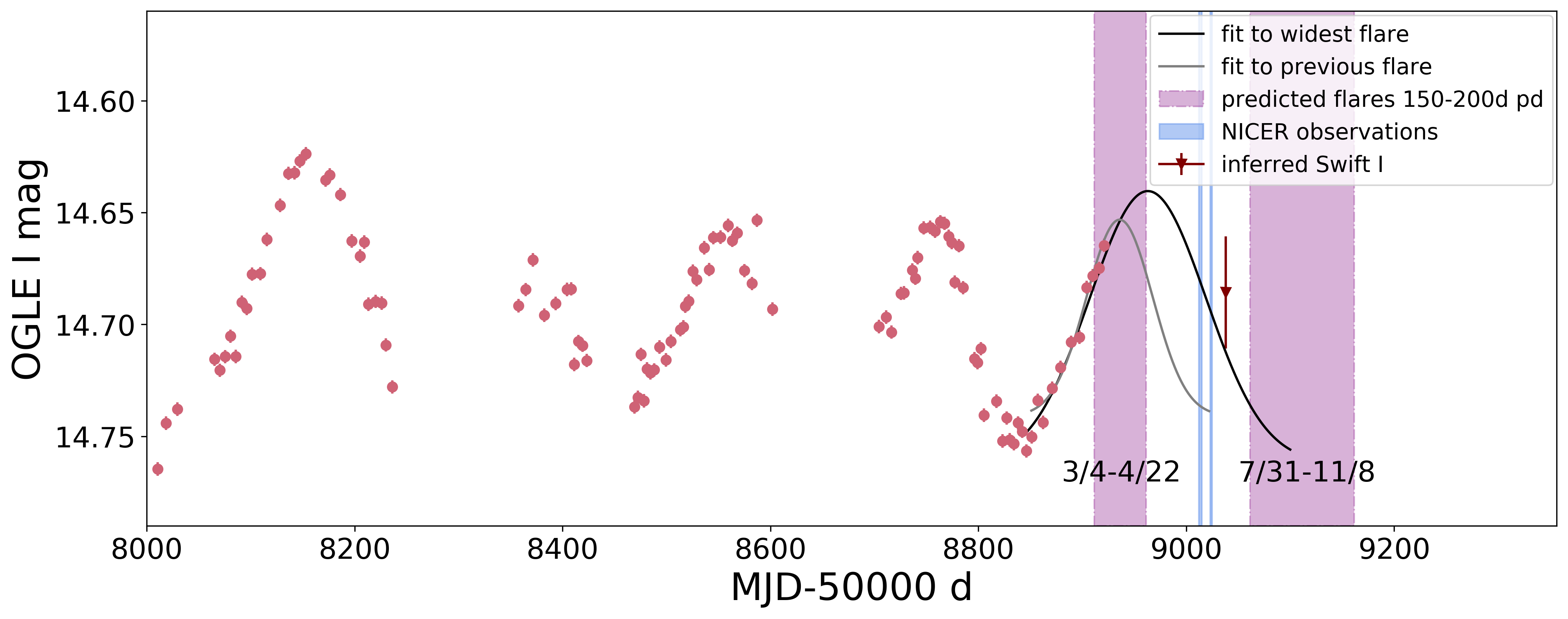}
    \caption{Final few peaks observed by OGLE, ending in March 2020. Purple shaded regions show the expected following two flare peaks, using periods between 150 and 200 days from the last peak (at 8761 days). The shaded regions imply that the X-ray outburst in June 2020 would not have coincided with an I band maximum. The V band \swift observations and $V-I_{int}$ plot yield the inferred I band value. The black Gaussian shows the fit to the widest flare in the light curve, showing an approximate limit on how close the X-ray outburst could have been to an I band flare. The grey curve is the Gaussian fit to the previous flare, shifted by $\sim$175 days.} 
    \label{fig:recent}
\end{figure}

Using the model of periodicity from the ten-year OGLE light curve, we attempted to extrapolate the optical behavior during the June 2020 X-ray outburst. If we predict the two peaks following the final full one, using periods between 150 and 200 days, we expect the optical data to be near a minimum in June, when the X-ray outburst was observed (Fig.~\ref{fig:recent}). To present an approximate limit, we shifted the Gaussian fit of the widest flare of the light curve to align with the flare rise from March. This fit also coincides with the 200 day spacing from the previous flare as well as the inferred I band point using \swift V band observations. The \nicer and \ero observations would not align with the flare maximum of such a wide flare. We also show the fit from the previous flare, shifted by about 175 days. Such a flare would then predict the \nicer observations to be right near the minimum in the optical, but is less consistent with the periodicity of $\sim$200 days in the final epoch of the light curve.

In the long-term X-ray LC (Fig.~\ref{fig:fulllcs}), we see corroborating evidence that the X-ray and optical peaks do not coincide. The first \swift data point shown (MJD 56968) implies that the system was in quiescence just a few days from the maximum observed during the entire OGLE LC.
Furthermore, the X-ray burst at $\sim$58220 occurred near an optical minimum.  
One possible explanation for the opposing optical and X-ray behavior could be the geometry of the system. An edge-on Be disc -- revealed by bluer-when-brighter behavior -- obscures the star, which is thus fainter when the disc grows \citep[][]{2011MNRAS.413.1600R}. At periastron, the system would become fainter (and redder, thanks to the relatively cool disk), as is potentially the case with this system. However, from the analysis of the V-I colour evolution, we found that \lxp gets redder when brighter, implying that we are not viewing the Be disc edge-on.

\section{SALT: Optical spectrum}
\label{sec:salt}

\begin{figure}
    \centering
    \includegraphics[width=1.0\columnwidth]{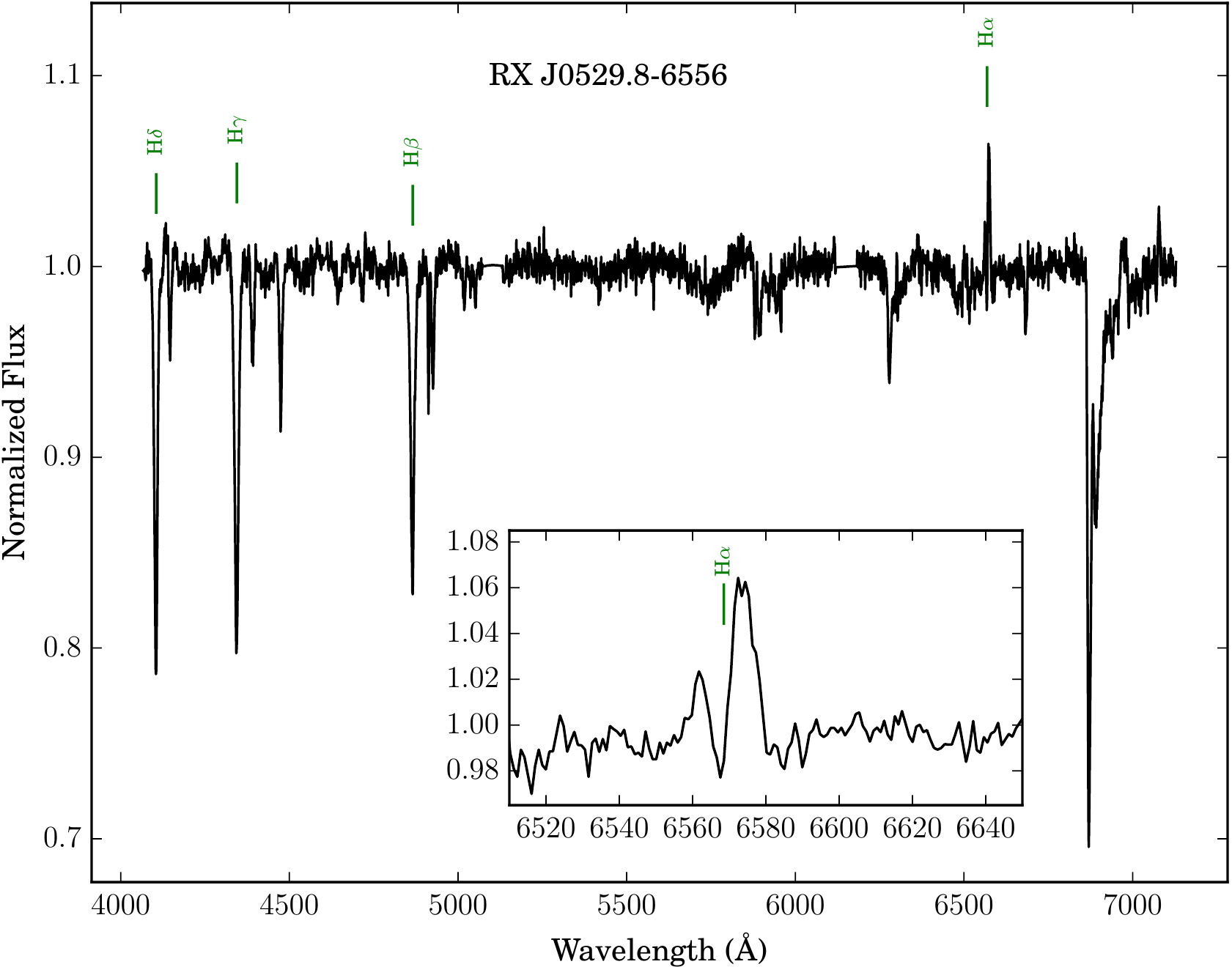}
    \caption{SALT spectrum of the optical counterpart of \lxp. The wavelength of Balmer lines in the LMC rest frame are marked. The inset shows a zoom-in of the \Halpha line region.}
    \label{fig:SALT}
\end{figure}

The optical counterpart of \rxj has been identified as a B0.5Ve star \citep{2002A&A...385..517N}.
Moreover, archival optical spectra \citep[from Nov. 1995, ][]{1997A&A...318..490H} show \Halpha in emission (EW = -0.9$\pm$0.2\,\AA), with all other Balmer lines in absorption.
\rosat observed \rxj 32 times over $\sim$3.5 years and detected only one strong X-ray outburst of the source.
The optical spectrum, which was obtained in 1995 after the \rosat observations, was most likely taken during an X-ray low state.
Following the June 2020 outburst of \lxp we obtained optical spectra with better resolution to study the \Halpha line properties.

The optical spectroscopy of \lxp was undertaken on 2020-09-11 (JD 2459103.5877) after the X-ray outburst using the Robert Stobie Spectrograph (RSS; \citealt{2003SPIE.4841.1463B}) on the Southern African Large Telescope (SALT; \citealt{2006SPIE.6267E..0ZB}) under the SALT transient followup program. 
The PG0900 VPH grating was used, which covered the spectral region 3920-7000 \AA\ at a resolution of 6.2 \AA. A single 1200 s exposure was obtained. The primary reductions, which comprise overscan and gain correction, bias subtraction and amplifier cross-talk corrections, were performed with the SALT pipeline \citep{2012ascl.soft07010C}. The remainder of the data reduction steps were carried out using \textsc{iraf\footnote{Image Reduction and Analysis Facility: iraf.noao.edu}} (identification of arc lines, background subtraction and extraction of the one-dimensional spectrum). The spectrum is dominated by deep Balmer absorption lines, with the  H$\alpha$ being the only line in emission (Fig.~\ref{fig:SALT}). The line exhibits an asymmetric double-peaked profile that is red-dominated ($V/R = 0.42\pm 0.05$). We note that the line morphology in the archival data could not be resolved due to limited spectral resolution.
The measured line equivalent width (EW) is -0.83$\pm$0.22, where uncertainties were estimated by following \citet{2016A&A...590A.122R}, and the largest contribution to the uncertainty is the continuum.
Finally, the peak separation ($\Delta V$) is $12.33\pm0.20$~\AA.

\section{Discussion}
\label{sec:disc}

\subsection{Pulse shape probes a super critical regime during 2020 outburst.}
\label{sec:supercritical}
From the discovery of the first pulsating X-ray systems it was clear that an anisotropic accretion mechanism was needed to explain the observed X-ray pulsations.
It was thus proposed that in highly magnetized compact stars accretion occurs through an accretion funnel that drives matter onto the magnetic poles of the compact star \citep{1973A&A....22..421B}. As a first attempt to explain the observed pulsations, it was suggested that the observed pulse profile can be a result of two components, a knife (or fan) and a pencil beam \citep{1973A&A....25..233G}. 
The term pencil beam was initially used to describe a low-luminosity accretion regime where the radiation escaped parallel to the magnetic field lines of the NS. Above a critical limit, the accretion rate becomes locally super-Eddington and a shock is formed above the NS surface, creating a thermal mound referred to as the accretion column. In this regime radiation escapes perpendicular to the magnetic field lines in a fan beam pattern \citep{1976MNRAS.175..395B}. 

Observationally, changes in the morphology of pulse profiles with luminosity are typical for X-ray pulsars \citep[e.g.,][]{1989ApJ...338..373P,2017MNRAS.466.2143D,2018A&A...614A..23K,2020MNRAS.493.5680J}. 
High luminosity is associated with double-peaked profiles and fan beam emission patterns, while sinusoidal pulse profiles are traditionally attributed to lower accretion rates and pencil beam emission patterns.
Moreover, pulse profiles are also energy dependent, with strong changes in their shape appearing around the cyclotron energies $E_{\rm cyc}$ \citep{2014A&A...564L...8S}. 
The cyclotron energy is expressed as $E_{\rm cyc}\approx11.57\times{B12}(1+z)^{-1}$ keV, where $B12$ is the $B$ field in units of $10^{12}$ G and $z$ is the gravitational redshift from the NS \citep[see][]{2007A&A...472..353S}.
For typical values of NS $B$ field these changes appear at energies above 10 keV.
However, there are cases where pulse profile transitions are evident within the 0.3-10.0 keV range \citep[e.g.,][]{2017MNRAS.470.1971V}. 
For \lxp, a double-peaked pulse profile as the one in Fig. \ref{fig:3panel}, is characteristic for a fan-beam emission, and thus a regime where the accretion column has been formed.

A rapid change in the pulse profile (for small change in $L_{\rm X}$) is not uncommon for X-ray pulsars \citep[e.g.,][]{2020MNRAS.493.5680J}. For example 4U\,1901+03 has shown sudden changes between six different pulse profile morphology between $L_{\rm X}$ of 0.7--1.2\ergs{37} \citep{2020MNRAS.493.5680J}. For 4U\,1901+03 an asymmetric pulse profile with a minor peak leading a major peak appeared in the brightest stages \citep{2020MNRAS.493.5680J}. Archival \rosat observations have shown a similar asymmetric pulse profile at $L_{\rm X}$ of 5\ergs{35}, however at that stage the major peak was leading in pulse phase \citep{2016A&A...594A..45R}. 
Similarly, shapes of pulse profiles at luminosity levels that differ by $\sim$100 (i.e., 1--100\ergs{37}) are also seen from Swift\,J0243.6+6124 \citep[][]{2018ApJ...863....9W}.
Detailed modeling of the pulse profile evolution is beyond the scope of the paper. However, our findings suggest that \lxp might be an intriguing case study for modeling the pulse profiles of the same systems at different luminosity levels \citep[e.g., with codes like Polestar; ][]{2017PASP..129l4201C}.

\subsection{Estimating the NS Magnetic Field Strength}
Apart from changes in pulse shape, the formation of the accretion column at a critical luminosity
 $L_{\rm crit}$ has been proposed to mark a pivot point in the spectral hardness evolution with $L_{\rm X}$ that is also seen in  observations \citep{2013A&A...551A...1R}.
Following \citet{2012A&A...544A.123B} this luminosity is given by \citep[see also][]{2015MNRAS.452.1601P}:
\begin{equation}
L_{\rm crit}=\left(\frac{B}{0.688\times10^{12}~G}\right)^{16/15}{\times}10^{37}~{\rm erg/s},
\label{eq1}
\end{equation}
that holds for typical parameters \citep[see eq. 32 of][for more details]{2012A&A...544A.123B}.
 
For the case of \lxp we found that during the decay of the 2020 outburst,
we see a decrease in luminosity, accompanied by an increase in hardness, shown by the higher HR in Fig.~\ref{fig:3panel} and the lower spectral photon index (Table \ref{tab:spectral}), seen in the \nicer spectra with the higher count statistics. The softer when brighter trend is related to the super-critical regime.
This trend is also consistent with the change in the pulse profile morphology. In particular, the transition between a profile having two pulses with comparable heights (i.e., epoch 2) and two asymmetric pulses  (i.e., epoch 1) is proposed to be a characteristic of the change to the super-critical regime \citep{2020MNRAS.493.5680J}.
Thus we can use the luminosity range between epochs 1 and 2 as an approximate limit for $L_{\rm crit}$. 
Since our spectral analysis is limited in the 0.3-8.0 keV band, a scaling factor of 2-3 should be used to convert to bolometric $L_{\rm X}$ \citep[see typical scaling factors for BeXRBs;][]{2016MNRAS.461L..97J,2020MNRAS.494.5350V}.
Then by using eq. \ref{eq1} we obtain a $B$ field on the order of $\sim0.7-2\times10^{12}$ G. 
Moreover, given that we do not see any pulse profile evolution within the \nicer energy range, this is consistent with a cyclotron energy outside the \nicer band and thus consistent with this B estimate.
Finally, this value is consistent with the $B$ field distribution of BeXRB pulsars in the MCs \citep[][]{2017RAA....17...59C}.

\subsection{A possible geometrical configuration}
Using the observed properties of \lxp we can compare with those of similar systems as well as theoretical predictions.
We speculate that the observational effects of \lxp could be consistent with the Be disc plane and orbital plane being misaligned \citep[e.g.,][]{2012ASPC..464..177O}. In that case, despite an approximately face-on Be disk, we would not necessarily see the Be disc broadening in its own plane (broadening being the usual cause of optical, red flares) at periastron. Furthermore, the misaligned planes could contribute to effects such as disc precession that lead to super-orbital optical trends.
The BeXRB pulsar SXP\,5.05 was reported to have orthogonal orbital and disc planes \citep[][]{2015MNRAS.447.2387C}.
Furthermore, in SXP\,5.05, the NS eclipses the Be star, allowing for characterization of the Be disc and binary solution.
Like \lxp, the system showed redder when brighter behavior, X-ray and optical peaks that do not coincide, and super-orbital variations. 
Misaligned planes are also associated with eccentric disks that lead to the brighter and longer so-called Type II outbursts \citep[particularly in systems with lower spin and orbital periods than \lxp,][]{2016A&A...586A..81H} as well as cycles of depletion and precession in the Be disc \citep{2014ApJ...790L..34M}. In turn, a precessing Be disc could be causing a beat period that interferes in optical data with the orbital periodicity. The precession can also be affected by the mass supply by the Be star as well as evolution and reconnection events of inner and outer parts of the decretion disc \citep[][]{2012ASPC..464..177O}. 
In that scenario, a non-steady supply of material of the Be star to the decretion disc can cause the disc to break into an inner and outer part. 
The two regions evolve independently, and with different precession motions. Once the parts connect again, the outer disc will rapidly change its position angle to that of the inner disk. 
Such factors could cause a time-dependent beat period and rapid changes in the periodicity of the optical light curve.
To constrain the system geometry, we would need to observe more \lxp X-ray flares for timing and spectral analysis, including the establishment of the orbital period. 
Our measurement of the \Halpha line, and its separation of the two peaks, yields a difference in velocity space of $\Delta{V}= 564 \pm$9 km/s.
Assuming a Keplerian Be disk, its radius is $4GM_{*}\sin^2{(i)}\Delta{V}^{-2}$, where $i$ is the disc inclination.
Following the spectral classification of the donor as a B0.5Ve star, a mass of 18-25 M$_{\sun}$ may be adopted \citep[e.g.][]{1981Ap&SS..80..353S}. The radius of the Be disc is thus on the order of $\sim$55$~\sin^2{i}$ R$_{\sun}$. 
Assuming a circular orbit with an orbital period of 150-200 d, the binary separation is on the order of 400 R$_{\sun}$. Given that the measurement of the \Halpha occurred after the X-ray outburst, the small Be disc size might indicate a disc truncation. Alternatively, a highly eccentric orbit would be required to allow for type I outbursts. 
Further optical monitoring could reveal more about the super-orbital trend, which we have yet to see repeat. 
Radial velocity measurements could also be used to find the orbital period. 
The evolution of optical emission lines over the course of at least one orbit would help to constrain the Be disc behavior. 

\section{Summary} 
We used \ero, \nicer, and \swift to characterize the 2020 X-ray outburst of \lxp. We studied the evolution of the spectral and temporal properties with \ero and \nicer, and determined that the pulse profile morphology dramatically changed while the flux only decreased by a third. Given the softer when brighter evolution, we argue that the system is close to the critical luminosity, which puts an upper limit on the magnetic field. We analysed the ten years of archival OGLE monitoring. Surprisingly, we see sharp transitions in flare periodicity, making the best model three epochs with periods of 149, 171, and 200 days. There may be an underlying super-orbital trend that we have yet to see repeat. Because of these changes, the orbital period remains unknown. Extending these bounds on the flare periodicity to June 2020 shows that there was no optical flare during the X-ray outburst. This lack of correspondence was conclusively demonstrated using archival \swift and \xmm monitoring taken within the window of OGLE observations. We propose that the unusual geometry of a misaligned Be disc and orbital plane may explain these results. Additional X-ray and optical light curves and spectra are necessary to further explain the behavior of \lxp in both wavelength regimes.

\section*{Data availability}

X-ray data are available through the High Energy Astrophysics Science Archive Research Center \url{heasarc.gsfc.nasa.gov}. \\
OGLE data are available through the OGLE XROM online portal \url{http://ogle.astrouw.edu.pl/ogle4/xrom/xrom.html}. Datasets and analysis scripts are available on reasonable request to the authors.\\
The eROSITA data are subject to an embargo period of 24 months from the end of eRASS1 cycle. Once the embargo expires the data will be available upon reasonable request to the corresponding author.
Other data underlying this article will be shared on reasonable request to the corresponding author.

\section*{Acknowledgements}
The authors would like to thank the anonymous referee for their comments and suggestions that helped improve the manuscript.

This research has made use of data and/or software provided by the High Energy Astrophysics Science Archive Research Center (HEASARC), which is a service of the Astrophysics Science Division at NASA/GSFC.
Based on observations using: \xmm, an ESA Science Mission with instruments and contributions directly funded by ESA Member states and the USA (NASA); \swift, a NASA mission with international participation. 

The OGLE project has received funding from the National Science Centre, Poland, grant MAESTRO 2014/14/A/ST9/00121 to AU.

This work has made use of data from \ero, the primary instrument aboard SRG, a joint Russian-German science mission supported by the Russian Space Agency (Roskosmos), in the interests of the Russian Academy of Sciences represented by its Space Research Institute (IKI), and the Deutsches Zentrum für Luft- und Raumfahrt (DLR). The SRG spacecraft was built by Lavochkin Association (NPOL) and its subcontractors, and is operated by NPOL with support from the Max Planck Institute for Extraterrestrial Physics (MPE). 
The development and construction of the \ero X-ray instrument was led by MPE, with contributions from the Dr. Karl Remeis Observatory Bamberg \& ECAP (FAU Erlangen-N{\"u}rnberg), the University of Hamburg Observatory, the Leibniz Institute for Astrophysics Potsdam (AIP), and the Institute for Astronomy and Astrophysics of the University of T{\"u}bingen, with the support of DLR and the Max Planck Society. The Argelander Institute for Astronomy of the University of Bonn and the Ludwig Maximilians University Munich also participated in the science preparation for \ero.
The \ero data shown here were processed using the eSASS software system developed by the German \ero consortium.

HT is funded by the Dorrit Hoffleit Undergraduate Research Scholarship at Yale. HT and GV are supported by NASA Grant Number 80NSSC21K0213, in response to NICER cycle 2 Guest Observer Program. 
GV is supported by NASA Grant Number  80NSSC20K0803, in response to XMM-Newton AO-18 Guest Observer Program. 
GV acknowledges support by NASA Grant number 80NSSC20K1107. NICER research at NRL is supported by NASA.




\bibliographystyle{mnras}
\bibliography{general}



\newpage
\appendix

\section{Estimating peaks of optical maxima}
\label{sec:flarefit}

\begin{figure}
    \centering
    \includegraphics[width=1.0\columnwidth]{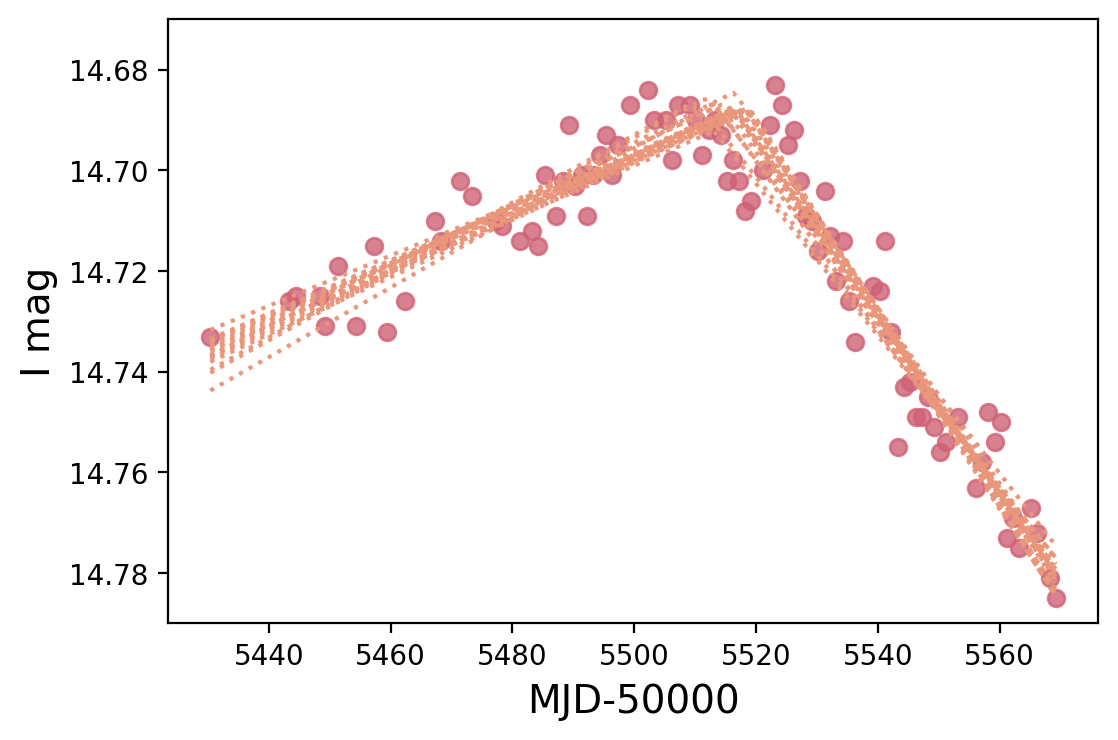}
    \caption{20 bootstrap best fits using the triangle model on the first optical peak. The resulting centers span ten days.}
    \label{fig:tribs}
\end{figure}

In the OGLE I band data we selected 13 intervals with clearly identifiable optical maxima. The optical peaks as well as the rise and decay of each peak were visible for each interval, thus enabling us to fit an analytical model and determine the epoch of each peak.    
There are several additional peaks in the light curve with insufficient data; since the peaks are often asymmetrical and irregularly-shaped (e.g. double-peaked), it does not make sense to use a model to extrapolate the peak center with only a few points. We were able to fit chunks of data, with varying confidence, as long as they showed at least a couple points on both sides near the peak. We used reasonable cutoffs for the start and end of each interval, including as much as possible without points from a neighboring optical peaks. We tested slight changes in the cutoff to make sure that the resulting best fit centers fell into our final uncertainty ranges.

Because of the irregular shapes, long-term underlying trend, and relatively low sampling, the choice of model is non-trivial. Furthermore, the main source of uncertainty on a given optical peak center is the systematic error associated with the model choice. For instance, the shapes are not fundamentally Gaussian. A Gaussian fit will give a low error on the center (a couple days), but choosing a different model can change the center result by several days.

We focused on three models: a Gaussian, a Gaussian + linear composite model (which allows for a continuum with nonzero slope), and a triangle model. For the Gaussian and Gaussian + linear models, we used the \texttt{LMFIT} package in Python \citep{2016ascl.soft06014N}. The different results from each model allowed us to quantify the systematic error. 
We also used bootstrapping to estimate the error within the model fit.

The OGLE LC shows a long-term trend on top of individual quasi-periodic cycles.
We tested two approaches to better model this characteristic: we fit the detrended data and used a composite model. Fitting the detrended data has a major drawback in that the window used for detrending affects the result. In addition, the results only change by a couple days at most. Our second model fit the optical peaks with both a Gaussian and a linear model, effectively (for 12 of them) fitting a tilted Gaussian: one with a continuum that is not parallel to the $x$-axis. For several optical peaks, such a model shifted the fit center by several days.

\begin{figure*}
    \centering
    \includegraphics[width=2.1\columnwidth]{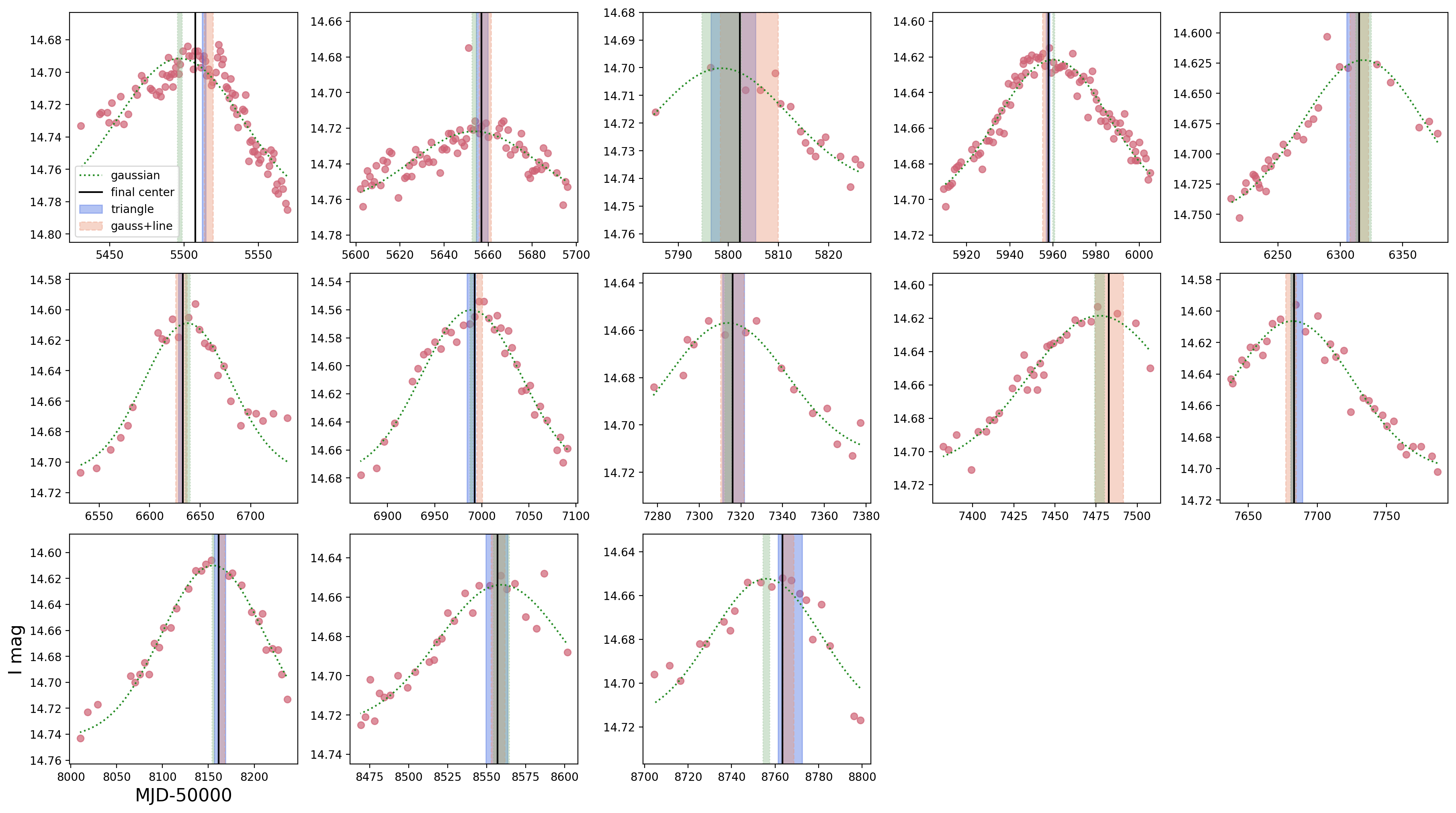}
    \caption{The 13 epochs with well defined optical peaks are show here.
    The Gaussian best-fit model is overlain on the I mag data points. Shading for each model represents the original center determinations +/- the standard deviation determined from 100 bootstrap iterations. The black lines indicate the final centers, chosen as the middle of the total span from the three methods.}
    \label{fig:fincenter}
\end{figure*}

The final model we tried was a triangle. The model is constructed with the four parameters being the slopes and intercepts of two intersecting lines; the intersection is the fit center. 
The model reasonably fits all the optical peaks.
The triangle model, however, is the most sensitive of the three models to the start and end cutoffs of the data included in a given chunk of data. 

Bootstrapping encompassed the sensitivities of each model to describe the error on each result. We used 100 bootstrap iterations for each interval (i.e., data set around each peak) and for each of the three models. We then took the standard deviation of each set of bootstrap results to be the error on the center for that model and flare. 20 bootstrap results using the triangle model are shown on the first peak of the light curve in Fig.~\ref{fig:tribs}. The Gaussian + linear model is quite sensitive in bootstrapping, so we removed outliers that were greater than 15 days from the original Gaussian + linear center. The method of bootstrapping was able to summarize the contributions of error from our other tests as well as the data sampling. The bootstrap results and the final centers used for periodicity analysis are shown in Fig.~\ref{fig:fincenter}. The lower and upper bounds for each center are the minimum and maximum of the three centers $\pm$ their bootstrap standard deviations. The center value is then the center of these bounds. The mean error is $\sim$seven days and the maximum is the error on the first optical peak: $\sim$12 days.




\bsp	
\label{lastpage}
\end{document}